\documentclass[twocolumn]{aastex701}

\begin{document}

\title{An infrared echo from a circumstellar disk in the hydrogen- and helium-poor SN\,2024aecx}

\correspondingauthor{Samaporn~Tinyanont}
\email{samaporn@narit.or.th}

\author[0000-0002-1481-4676]{Samaporn~Tinyanont}
\affiliation{National Astronomical Research Institute of Thailand, 260 Moo 4, Donkaew, Maerim, Chiang Mai, 50180, Thailand}
\email{samaporn@narit.or.th}

\author[0000-0002-7593-2748]{Kittipong~Wangnok}
\affiliation{National Astronomical Research Institute of Thailand, 260 Moo 4, Donkaew, Maerim, Chiang Mai, 50180, Thailand}
\affiliation{School of Physics, Institute of Science, Suranaree University of Technology, Nakhon Ratchasima, 30000, Thailand}
\email{kittipong@narit.or.th}

\author[0000-0003-0123-0062]{Jennifer E. Andrews}
\affiliation{Gemini Observatory/NSF's NOIRLab, 670 N. A'ohoku Place, Hilo, HI 96720, USA}
\email{jennifer.andrews@noirlab.edu}

\author[0000-0002-2445-5275]{Ryan~J.~Foley}
\affiliation{Department of Astronomy and Astrophysics, University of California, Santa Cruz, CA 95064, USA}
\email{foley@ucsc.edu}

\author[0009-0005-9062-9471]{Methawee~Kaewmookda}
\affiliation{National Astronomical Research Institute of Thailand, 260 Moo 4, Donkaew, Maerim, Chiang Mai, 50180, Thailand}
\email{methawee@narit.or.th}

\author[0000-0001-5754-4007]{Jacob~E.~Jencson}
\affiliation{IPAC, Mail Code 100-22, Caltech, 1200 E. California Blvd. Pasadena, CA 91125, USA}
\email{jjencson@ipac.caltech.edu}

\author[0000-0002-4410-5387]{Armin~Rest}
\affiliation{Space Telescope Science Institute, Baltimore, MD 21218, USA}
\affiliation{Department of Physics and Astronomy, Johns Hopkins University, Baltimore, MD 21218, USA}
\email{rest@stsci.edu}

\author[0000-0002-4449-9152]{Katie~Auchettl}
\affiliation{Department of Astronomy and Astrophysics, University of California, Santa Cruz, CA 95064, USA}
\affiliation{School of Physics, The University of Melbourne, Parkville, VIC, Australia}
\email{katie.auchettl@unimelb.edu.au}

\author[0000-0002-4924-444X]{K.~Azalee~Bostroem}
\affiliation{Steward Observatory, University of Arizona, 933 North Cherry Avenue, Tucson, AZ 85721-0065, USA}
\email{abostroem@gmail.com}

\author[0000-0003-4263-2228]{David~A.~Coulter}
\affiliation{Space Telescope Science Institute, Baltimore, MD 21218, USA}
\affiliation{Physics and Astronomy Department, Johns Hopkins University, Baltimore, MD 21218, USA}
\email{dcoulter@stsci.edu}

\author[0000-0002-9099-4613]{Poemwai~Chainakun}
\affiliation{School of Physics, Institute of Science, Suranaree University of Technology, Nakhon Ratchasima, 30000, Thailand}
\email{pchainakun@sut.ac.th}

\author[0000-0002-7706-5668]{Ryan~Chornock}
\affiliation{Department of Astronomy, University of California, Berkeley, CA 94720-3411, USA} 
\affiliation{Berkeley Center for Multi-messenger Research on Astrophysical Transients and Outreach (Multi-RAPTOR), University of California, Berkeley, CA 94720-3411, USA}
\email{chornock@berkeley.edu}

\author[0000-0002-5680-4660]{Kyle~W.~Davis}
\affiliation{Department of Astronomy and Astrophysics, University of California, Santa Cruz, CA 95064, USA}
\email{kywdavis@ucsc.edu}

\author[0000-0003-2238-1572]{Ori D. Fox}
\affiliation{Space Telescope Science Institute, Baltimore, MD 21218, USA}
\email{ofox@stsci.edu}

\author[0000-0002-1296-6887]{Llu\'is Galbany}
\affiliation{Institute of Space Sciences (ICE-CSIC), Campus UAB, Carrer de Can Magrans, s/n, E-08193 Barcelona, Spain}
\affiliation{Institut d'Estudis Espacials de Catalunya (IEEC), 08860 Castelldefels (Barcelona), Spain}
\email{l.g@csic.es}

\author[0000-0003-2824-3875]{T. R. Geballe}
\affiliation{Gemini Observatory/NSF's NOIRLab, 670 N. A'ohoku Place, Hilo, HI 96720, USA}
\email{tgeballe@gmail.com}

\author[0000-0002-9454-1742]{Brian~Hsu}
\affiliation{Steward Observatory, University of Arizona, 933 North Cherry Avenue, Tucson, AZ 85721-0065, USA}
\email{bhsu@arizona.edu}

\author[0000-0002-3934-2644]{Wynn~V.~Jacobson-Gal\'an}
\altaffiliation{NASA Hubble Fellow}
\affiliation{Division of Physics, Mathematics and Astronomy, California Institute of Technology, 1200 E. California Blvd., Pasadena, CA 91125, USA}
\email{wynnjg@caltech.edu}

\author[0000-0001-8738-6011]{Saurabh~W.~Jha}
\affiliation{Department of Physics and Astronomy, Rutgers, The State University of New Jersey, Piscataway, NJ 08854, USA}
\email{saurabh@physics.rutgers.edu}

\author[0009-0005-1871-7856]{Ravjit Kaur}
\affiliation{Department of Astronomy and Astrophysics, University of California, Santa Cruz, CA 95064, USA}
\email{ravkaur@ucsc.edu}

\author[0000-0002-5619-4938]{Mansi~M.~Kasliwal}
\affiliation{Division of Physics, Mathematics and Astronomy, California Institute of Technology, 1200 E. California Blvd., Pasadena, CA 91125, USA}
\email{mansi@astro.caltech.edu}

\author[0000-0003-0778-0321]{Ryan~M.~Lau}
\affiliation{IPAC, Mail Code 100-22, Caltech, 1200 E. California Blvd. Pasadena, CA 91125, USA}
\email{ryanlau@ipac.caltech.edu}

\author[0000-0002-2249-0595]{Natalie~LeBaron}
\affiliation{Department of Astronomy, University of California, Berkeley, CA 94720-3411, USA}
\affiliation{Berkeley Center for Multi-messenger Research on Astrophysical Transients and Outreach (Multi-RAPTOR), University of California, Berkeley, CA 94720-3411, USA}
\email{nlebaron@berkeley.edu}

\author[0000-0003-4768-7586]{Rafaella Margutti} 
\affiliation{Department of Astronomy, University of California, Berkeley, CA 94720-3411, USA}
\affiliation{Berkeley Center for Multi-messenger Research on Astrophysical Transients and Outreach (Multi-RAPTOR), University of California, Berkeley, CA 94720-3411, USA}
\email{rmargutti@berkeley.edu}

\author[0000-0001-7488-4337]{Seong~Hyun~Park}
\affiliation{Department of Physics and Astronomy, Seoul National University, Gwanak-ro 1, Gwanak-gu, Seoul 08826, South Korea}
\email{rogersh0125@snu.ac.kr}

\author[0000-0002-0744-0047]{Jeniveve~Pearson}
\affiliation{Steward Observatory, University of Arizona, 933 North Cherry Avenue, Tucson, AZ 85721-0065, USA}
\email{jenivevepearson@arizona.edu}

\author[0000-0001-6806-0673]{Anthony~L.~Piro}
\affiliation{The Observatories of the Carnegie Institution for Science, 813 Santa Barbara St., Pasadena, CA 91101, USA}
\email{piro@carnegiescience.edu}

\author[0000-0003-4175-4960]{Conor~L.~Ransome}
\affiliation{Steward Observatory, University of Arizona, 933 North Cherry Avenue, Tucson, AZ 85721-0065, USA}
\email{cransome@arizona.edu}

\author[0000-0002-7352-7845]{Aravind~P.~Ravi}
\affiliation{Department of Physics and Astronomy, University of California, Davis, 1 Shields Avenue, Davis, CA 95616-5270, USA}
\email{apazhayathravi@ucdavis.edu}

\author[0000-0003-3643-839X]{Jeonghee Rho}
\affiliation{SETI Institute, 189 Bernardo Ave., Ste. 200, Mountain View, CA 94043, USA} 
\affiliation{Department of Physics and Astronomy, Seoul National University, Gwanak-ro 1, Gwanak-gu, Seoul, 08826, South Korea}
\email{jrho@seti.org}

\author[0000-0002-7559-315X]{César Rojas-Bravo}
\affiliation{School of Astronomy and Space Science, University of Chinese Academy of Sciences, Beijing 100049, People’s Republic of China}
\affiliation{National Astronomical Observatories, Chinese Academy of Sciences, Beijing 100101, People’s Republic of China}
\email{cesar.rojasbravo@ucas.ac.cn}

\author[0000-0003-4725-4481]{Sam~Rose}
\affiliation{Division of Physics, Mathematics and Astronomy, California Institute of Technology, 1200 E. California Blvd., Pasadena, CA 91125, USA}
\email{srose@caltech.edu}

\author[0000-0003-4102-380X]{David~J.~Sand}
\affiliation{Steward Observatory, University of Arizona, 933 North Cherry Avenue, Tucson, AZ 85721-0065, USA}
\email{dsand@arizona.edu}

\author[0000-0001-5510-2424]{Nathan~Smith}
\affiliation{Steward Observatory, University of Arizona, 933 North Cherry Avenue, Tucson, AZ 85721-0065, USA}
\email{nathans@as.arizona.edu}

\author[0000-0002-4022-1874]{Manisha~Shrestha}
\affiliation{School of Physics and Astronomy, Monash University, Clayton, Australia}
\email{manisha.shrestha@monash.edu}

\author[0000-0001-8073-8731]{Bhagya~M.~Subrayan}
\affiliation{Steward Observatory, University of Arizona, 933 North Cherry Avenue, Tucson, AZ 85721-0065, USA}
\email{bsubrayan@arizona.edu}

\author[0000-0001-8818-0795]{Stefano Valenti}
\affiliation{Department of Physics and Astronomy, University of California, Davis, 1 Shields Avenue, Davis, CA 95616-5270, USA}
\email{stfn.valenti@gmail.com}

\begin{abstract}
We present near-infrared (NIR) spectroscopy of the hydrogen- and helium-poor (Type Ic) supernova (SN) 2024aecx, which displays a strong NIR excess emerging 32 days post peak. 
SN\,2024aecx is a peculiar SN Ic that exhibited luminous shock-cooling emission at early times, suggestive of close-in circumstellar medium (CSM), unexpected for this class of SNe. 
Its early NIR spectra are typical for a SN Ic but with strong \ion{C}{1} absorption features.
By $\sim$32 days post peak, the spectra show a strong NIR excess, while maintaining normal optical colors, unprecedented for SNe Ic. 
We find that the NIR excess is well fit with a single-temperature, optically thin dust model with declining temperature, increasing mass, and roughly constant luminosity over time.
The NIR excess appears too promptly for dust to have formed in the SN ejecta, indicating an IR echo from pre-existing dust in the CSM. 
The IR echo is likely powered by the relatively slowly evolving SN peak light, and not the brief shock cooling emission, as the latter requires unrealistically high CSM densities to explain the observed dust mass.
We consider different potential CSM geometries and find that a thick face-on disk with an inner edge of around $5\times 10^{16} \rm \ cm$ can best explain the dust mass and temperature evolution.
In this scenario, the SN shock should start interacting with this CSM $440\pm200$ days post explosion.
CSM around SN Ic is rare, and follow-up observations of SN\,2024aecx will probe the mass-loss process responsible for removing hydrogen and helium from their progenitor star. 
\end{abstract}


\section{Introduction} \label{sec:intro}
Massive stars ($M_{\rm ZAMS} \gtrsim 8\ M_\odot$) lose a significant amount of mass prior to their demise as core-collapse supernovae (CCSNe). 
About a third of CCSNe \citep[e.g.,][]{Smith2011rate, Shivvers2017} show little to no hydrogen in their spectra (Type IIb and Ib, respectively), or even no helium (Type Ic; \citealp[see e.g.,][for reviews of SN spectral classification]{Filippenko1997, Gal-Yam2017}).
The mass-loss process responsible for removing hydrogen and/or helium from the progenitor to these stripped-envelope supernovae (SESNe) remains debated \citep[see, e.g.,][for a review]{Smith2014}, but binary interaction is likely the main culprit \citep[e.g.,][]{Eldridge2013, Yoon2017,Zapartas2025}.
In binary mass transfer, the majority of mass loss happens when the primary evolves off the main sequence, long before its death \citep[e.g.,][]{Yoon2017}. 
Thus, any circumstellar medium (CSM) resulting from non-conservative mass loss is expected to be far away from the progenitor star when it explodes.

Yet, in a small number of SESNe, the hydrogen-rich circumstellar medium (CSM) remains close to the progenitor such that we observe the SN shock ($v_{\rm shock}\approx 10{,}000\ \rm km\,s^{-1}$) crashing into it within months. 
Examples include SNe\,2001em \citep{Chugai2006, Chandra2020}, 2004dk \citep{Mauerhan2018, Pooley2019, Balasubramanian2021}, 2018ijp \citep{Tartaglia2021}, 2019oys \citep{Sollerman2020, Sfaradi2024}, 2019tsf \citep{Sollerman2020, Zenati2025}, 2019yvr \citep{Kilpatrick2021, Ferrari2024}, and 2021efd \citep{Pyykkinen2025}.
The best-observed and most representative case is SN\,2014C \citep{Milisavljevic2015, Margutti2017, Tinyanont2019, Brethauer2022, Thomas2022,  Zhai2025, Tinyanont2025}.
It exploded as a normal SN Ib and started to develop strong and persistent intermediate-width ($\sim1000\ \rm km\,s^{-1}$) H$\alpha$ from a few weeks \citep{Milisavljevic2015, Zhai2025} until at least 3800 days post explosion \citep{Tinyanont2025}. 
Infrared (IR) observations of SN\,2014C show strong dust emission initially ($\lesssim 5 \ \rm yr$) from pre-existing CSM dust heated by the SN shock \citep{Tinyanont2019}, to recent James Webb Space Telescope (JWST) observations revealing new dust formation in the cold dense shell (CDS) between the forward and reverse shocks \citep{Tinyanont2025}. 
Panchromatic observations from the radio to X-ray point to an inclined disk-like CSM.
Such a CSM close to the progenitor at death is not expected in the simple Case B binary mass transfer (when the primary evolves off of the main sequence), pointing to additional mass loss that occurs in a small fraction of SESNe.

In SNe with CSM, in addition to shock interactions, the SN light heats pre-existing circumstellar dust to a temperature that is dependent on SN luminosity, grain properties, and distance from the SN.
Close-in grains can be heated past their sublimation temperature and destroyed.
Dust that survives subsequently cools by radiating in the IR with the emission from different parts of the CSM arriving at different times, producing an IR echo \citep[e.g.,][]{Bode1980, Dwek1983, Graham1983}.
The evolution of the IR echo puts a strong constraint on the CSM geometry.
We note that these are distinct from light echoes from SN light scattering off of dust grains, which are prominent in the blue optical.
IR echoes can probe the CSM many (up to $\sim$30) times further than shock interactions at the same epoch (since $c \approx 30 v_{\rm shock}$, but the distance probed by the echo depends on the CSM geometry).
The trade-off is that dust much further out only gets heated to a few hundred kelvins, not easily detectable in the near-infrared (NIR) from the ground.
{The rise time of the IR echo is $r_{\rm inner, subl}/c$, where $r_{\rm inner, subl}$ is either the inner edge of the CSM or the dust sublimation radius, whichever is smaller. In a typical SN, $r_{\rm subl} \approx 10^{16} \ \rm cm$, leading to an echo rise time of only a few days, in a stark contrast with the expected dust formation timescales in the ejecta of several hundred days \citep[see e.g.,][for a review]{Sarangi2018}.
}
IR echoes from CSM dust have been observed in several types of CCSNe; examples include SNe\,1979C, 1980K \citep[II-L;][]{Bode1980, Dwek1983, Sugerman2012}, 1982G \citep[II;][]{Graham1983}, 1987A \citep[II-pec; e.g.,][]{Crotts1988, Rank1988}, 1993J \citep[IIb;][]{Lewis1994}, 2002hh \citep[II-P;][]{Pozzo2006, Meikle2006}, 2006jc \citep[Ibn;][]{Mattila2008}, 2010jl \citep[IIn;][]{Andrews2011, Dwek2021}, 2015da \citep[IIn;][]{Tartaglia2020}, and 2023xgo \citep[Ibn/Icn;][]{Yamanaka2025}.
These observations constrain the mass and distance of the CSM known from other interaction signatures, and constrain the luminosity of the unobserved shock breakout emission in some cases.
Although IR echoes have been suggested as a way to probe CSM around normal SNe Ib/c \citep{Chugai2006}, which would probe their elusive mass-loss mechanism, {no IR excess consistent with IR echoes have been reported in the literature for SNe Ib/c.}

SN\,2024aecx was discovered by the Asteroid Terrestrial-impact Last Alert System (ATLAS; \citealp{Tonry2018, Smith2020}) in NGC\,3521 on 2024 December 16 at 13:22 UT \citep[][UT dates and times used hereafter]{Stevance2024, Tonry2024}, and reported to the Transient Name Server (TNS). 
The Tully--Fisher distance modulus to the host galaxy is $30.70 \pm 0.43$, corresponding to a distance of $13.8\pm 2.7$~Mpc \citep{Sorce2014} and a redshift of $z = 0.002665$ \citep{Springob2005}.
The SN was notable for its early discovery, with a deep pre-explosion limit taken less than one day prior by the Zwicky Transient Facility (ZTF; photometry obtained via the forced photometry service; \citealp{Masci2023}), putting the explosion epoch at a Modified Julian Date (MJD) of 60660.04, {taken as the mid point between the last deep non detection and the first detection}. 
Rapid photometric follow-up observations by Kinder \citep{Chen2024} and further observations by ATLAS show blue and rapidly declining light curves, reminiscent of shock cooling emission typically seen in SNe IIb \citep{Perez-Fournon2024}, {and Ca-strong SNe Ib/c \citep[e.g.,][]{De2018, Jacobson-Galan2020, Irani2024, Yadavalli2024}.}
Early spectra taken on 2024 December 17 during the shock-cooling phase are largely featureless, leading to a preliminary Type IIb classification based on its light curve \citep{ Andrews2024a, Hinkle2024}. 
However, a Gemini spectrum taken two days later on 2024 December 19 at the start of the main nickel-powered SN peak better matches early-time spectra of SNe Ic, in particular SN\,1994I \citep{Andrews2024b}.
Its proximity and early observations led to successful Director's Discretionary Time (DDT) programs with JWST to follow up on this object in the NIR and mid-IR (PIDs 9233, 9258, PI Shahbandeh), while \citet{Zou2025} and \citet{Xi2025} presented optical observations of this SN. 

In this paper, we present our follow-up ground-based NIR spectroscopy of SN\,2024aecx showing a rapid onset of an IR excess between 12 and 32 days post peak.
We describe the observations in Section~\ref{sec:obs} and discuss host extinction in \ref{sec:extinction}. 
We discuss the spectral line evolution in Section~\ref{sec:spec_evo}; and the NIR continuum evolution and the corresponding dust properties in Section~\ref{sec:ir_excess}.
We provide conclusions in Section~\ref{sec:conclusion}.

\section{Observations and data reduction} \label{sec:obs}

\begin{figure*}
    \centering
    \includegraphics[width=\linewidth]{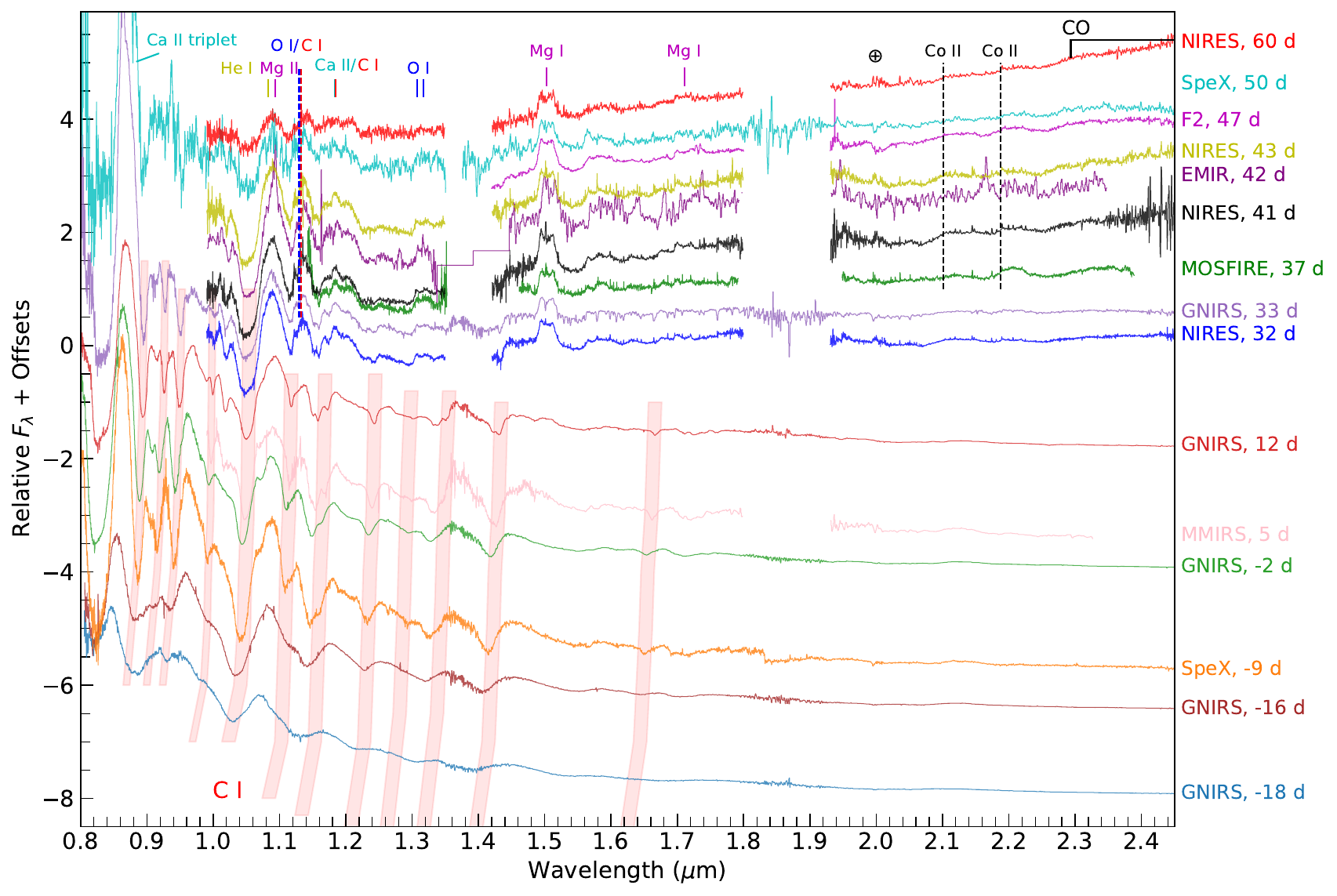}
    \caption{NIR spectra of SN\,2024aecx from $-18$ to 60 days from peak. 
    Prominent spectral features are marked. 
    Early spectra are dominated by a hot continuum with prominent \ion{C}{1} absorptions, marked by the shaded regions.  
    Between 12 and 32 days, a strong NIR continuum emerges and strengthens toward the end of the spectral sequence shown here. 
    Spectra in this phase show boxy \ion{Mg}{1} 1.5033 $\mu$m. 
    Comparatively weak CO first overtone is detected starting potentially from 37 days.
    The \ion{Ca}{2} triplet remains strong at all phases.
    }
    \label{fig:ir_spec}
\end{figure*}

Table~\ref{tab:spec_lob} lists all spectroscopic observations. 
We performed NIR spectroscopy using several echellette spectrometers on different telescopes, which cover 1--2.4 $\mu$m simultaneously. 
All observations were performed with an ABBA dithering pattern along the slit to sample the sky background. 
An A0V star (usually HIP54815) was observed immediately before the SN to provide flux and telluric calibration. 
Wavelength calibration is performed using the bright NIR sky lines. 
Figure~\ref{fig:ir_spec} shows all NIR spectra of SN\,2024aecx {and the spectra (uncorrected for reddening) are available on WISeREP \citep{Yaron2012}}.

\begin{table*}
    \caption{Log of spectroscopic observations}
    \centering
    \begin{tabular}{llllll}
    \toprule
    UT Date & MJD & Days from peak & Telescope/Instrument & $R$ & Exp. Time (s) \\ \hline
2024-12-18  & 60662 & -18 & Gemini-N/GNIRS &1200& 8$\times$180\\
2024-12-20  & 60664 & -16 & Gemini-N/GNIRS &1200&8$\times$180\\
2024-12-27  & 60671 & -9 & IRTF/SpeX  & 750& 12$\times$200\\
2025-01-03  & 60678 & -2 & Gemini-N/GNIRS &1200&8$\times$180\\
2025-01-11  & 60686 & 5  & MMT/MMIRS &2400 ($zJ$) 1400 ($HK$)& 720 ($zJ$) 720 ($HK$)\\
2025-01-18  & 60693 & 12 & Gemini-N/GNIRS &1200&8$\times$180\\
2025-02-07  & 60713 & 32 & Keck/NIRES & 2700 & 4$\times$200 \\ 
2025-02-08  & 60714 & 33 & Gemini-N/GNIRS &1200&8$\times$180\\
2025-02-12  & 60718 & 37 & Keck/MOSFIRE  & 2300 & 400 ($JH$) 480 ($K$) \\
2025-02-16  & 60722 & 41 & Keck/NIRES & 2700 & 2$\times$300 \\
2025-02-17  & 60723 & 42 & GTC/EMIR & 987 & 8$\times$120 ($YJ$) 8$\times$120 ($HK$) \\ %
2025-02-18  & 60724 & 43 & Keck/NIRES & 2700 & 4$\times$250 \\
2025-02-22  & 60728 & 47 & Gemini-S/F2 &600& 8$\times$120\\
2025-02-25  & 60731 & 50 & IRTF/SpeX  & 750 & 14$\times$300\\
2025-03-07  & 60741 & 60 & Keck/NIRES & 2700 & 4$\times$150 \\ 
         \hline
    \end{tabular}
    \label{tab:spec_lob}
\end{table*}

We observed SN\,2024aecx with the Near-InfraRed Echellette Spectrometer \citep[NIRES;][]{Wilson2004} on the 10~m Keck~II telescope, primarily as part of the Keck Infrared Transient Survey (KITS; \citealp{Tinyanont2024}).
NIRES data are reduced using \texttt{pypeit} \citep{prochaska2020, pypeit2020} following the procedure outlined by \citet{Tinyanont2024}.
The pipeline automatically performs flat-fielding, background subtraction, and source detection and extraction. 
The science spectra were then flux calibrated, coadded, and corrected for telluric absorption, using the aforementioned A0\,V star observation.
We also performed photometry on the acquisition images of the SN from the slit-viewing camera of NIRES, which has a $K'$ band filter.  
We used a custom pipeline to perform flat-field correction and background subtraction. 
We used stars observed by the UKIRT (United Kingdom InfraRed Telescope) Infrared Deep Sky Survey (UKIDSS; \citealp{Lawrence2007}) to perform photometric calibration. 
The SN photometry is used to provide the absolute flux calibration for the spectra. 
Due to the small number of comparison stars in the small field of view, resulting photometry is only good to about $\pm 0.1$ mag (10\%).
We note that this is still better than the distance uncertainty to the SN host; thus, uncertainties on results relying on the absolute flux scale are still dominated by the distance uncertainty. 

Multiple spectra were obtained with GNIRS on the 8-m Gemini North telescope \citep{Elias2006}. The data were taken in the cross-dispersed mode using the 32 l/mm grating and the 0\farcs675 slit.  
Data were reduced using the Gemini IRAF package, and were telluric and flux calibrated order by order using \texttt{xtellcor} \citep{vacca03} in the \texttt{spextool} \citep{cushing04} package with an A0V standard star taken at the same time.
Since GNIRS covers the optical $z$ band, we use photometry from \citet{Zou2025} for absolute flux calibration. 
We note that the GNIRS spectrum from 2025~February~08 calibrated using the $z$-band photometry and NIRES spectrum from February~07 calibrated using $K$ band agree within the uncertainties, to a few percent. 

We also obtained two spectra using SpeX on the NASA InfraRed Telescope Facility (IRTF), in the short cross-dispersed mode paired with the 0\farcs8 slit, providing simultaneous wavelength coverage from 0.7 to 2.55~$\mu$m at a resolving power of $R\sim$750.
Data were reduced using \texttt{spextool} \citep{cushing04} and \texttt{xtellcor} \citep{vacca03}.

In addition to echellette spectra, we obtained one spectrum using the Multi-Object Spectrometer for Infra-Red Exploration \citep[MOSFIRE;][]{mclean2012} on Keck~I in the long-slit configuration with a 1\farcs0 slit on 2025~February~12. 
The observations were done in the $J$, $H$, and $K$ bands (no $Y$ band). 
The data were reduced and calibrated with \texttt{pypeit} \citep{prochaska2020, pypeit2020}.

We also obtained a spectrum using the FLAMINGOS-2 (F2) spectrograph on the 8-m Gemini South telescope \citep{E2004, E2012} on 2025~February~22, with exposure times of 8 × 120 seconds, using the \textit{HK} grating setup covering 1.4–2.5 $\mu$m.
The data were reduced using Python-based IRAF scripts and were also compared with results from \texttt{pypeit}. 
We also used the A0V star HIP\,54815 for telluric calibration, but the star was observed on a different night, resulting in a substantial residual telluric absorption remaining in the corrected spectrum.

{We obtained one epoch of NIR spectra (\textit{zJ} and \textit{HK}) with the MMT and Magellan Infrared Spectrograph \citep[MMIRS;][]{mmirs} on the 6.5-m MMT located on Mt.\ Hopkins in Arizona, USA on 2025~Jan~11 at 11:07:57.282 UT (60686.46 MJD). The data were reduced using the MMIRS pipeline written in \texttt{IDL} \citep{mmirspipe}. Telluric and flux calibration were also performed using \texttt{xtellcor} \citep{vacca03} using a standard A0V star observed at a similar time and airmass.}

\begin{figure}
    \centering
    \includegraphics[width=\linewidth]{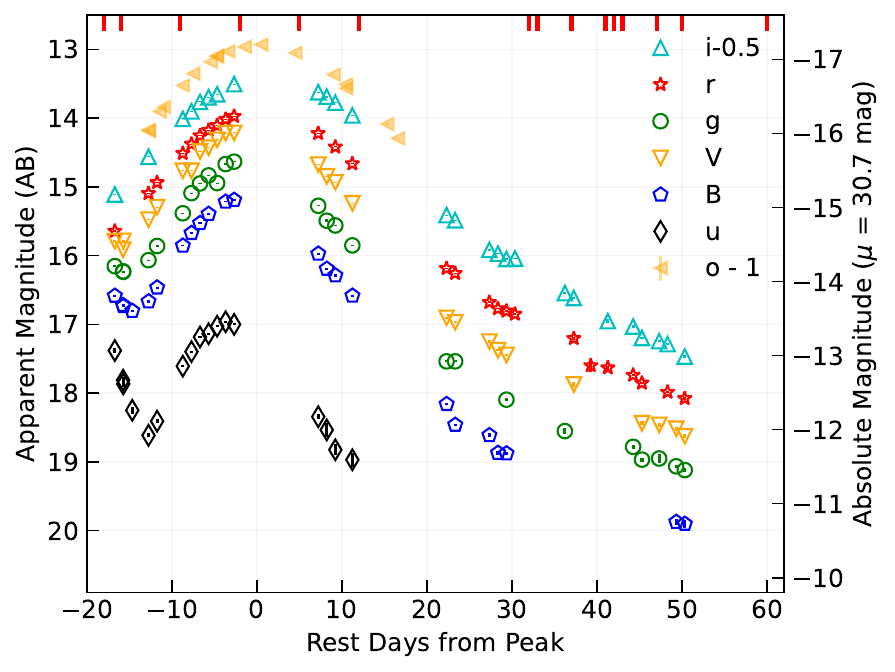}
    \caption{Optical light curves of SN\,2024aecx in the $uBVgri$ bands from Swope and the ATLAS $o$ band. The magnitudes are in the AB system. The photometry has not been corrected for extinction. {Red ticks on the top of the plot mark epochs for our NIR spectra.}}
    \label{fig:LC}
\end{figure}

{To find a reference peak epoch, we fit a third-order polynomial to the public ATLAS $o$-band photometry (shown in Figure~\ref{fig:LC}) and find the peak epoch of MJD 60680.26.}
We obtained optical photometry of SN\,2024aecx in the $uBVgri$ bands using the Direct 4k$\times$4k camera on the 1-m Henrietta Swope telescope at Las Campanas Observatory, Chile, from 2024 December 19 to 2025 April 14. The full data reduction process is described in \citet{Kilpatrick2018}, and the light curves are shown in Figure~\ref{fig:LC}. The photometry is in the natural system, consistent with the Carnegie Supernova Project-I \citep[CSP-I; e.g.,][]{Stritzinger2018phot}. {Further discussions on the optical light curve will be provided in Andrews et al. (2026, in preparation).}

\begin{figure*}
    \centering
    \includegraphics[width=\linewidth]{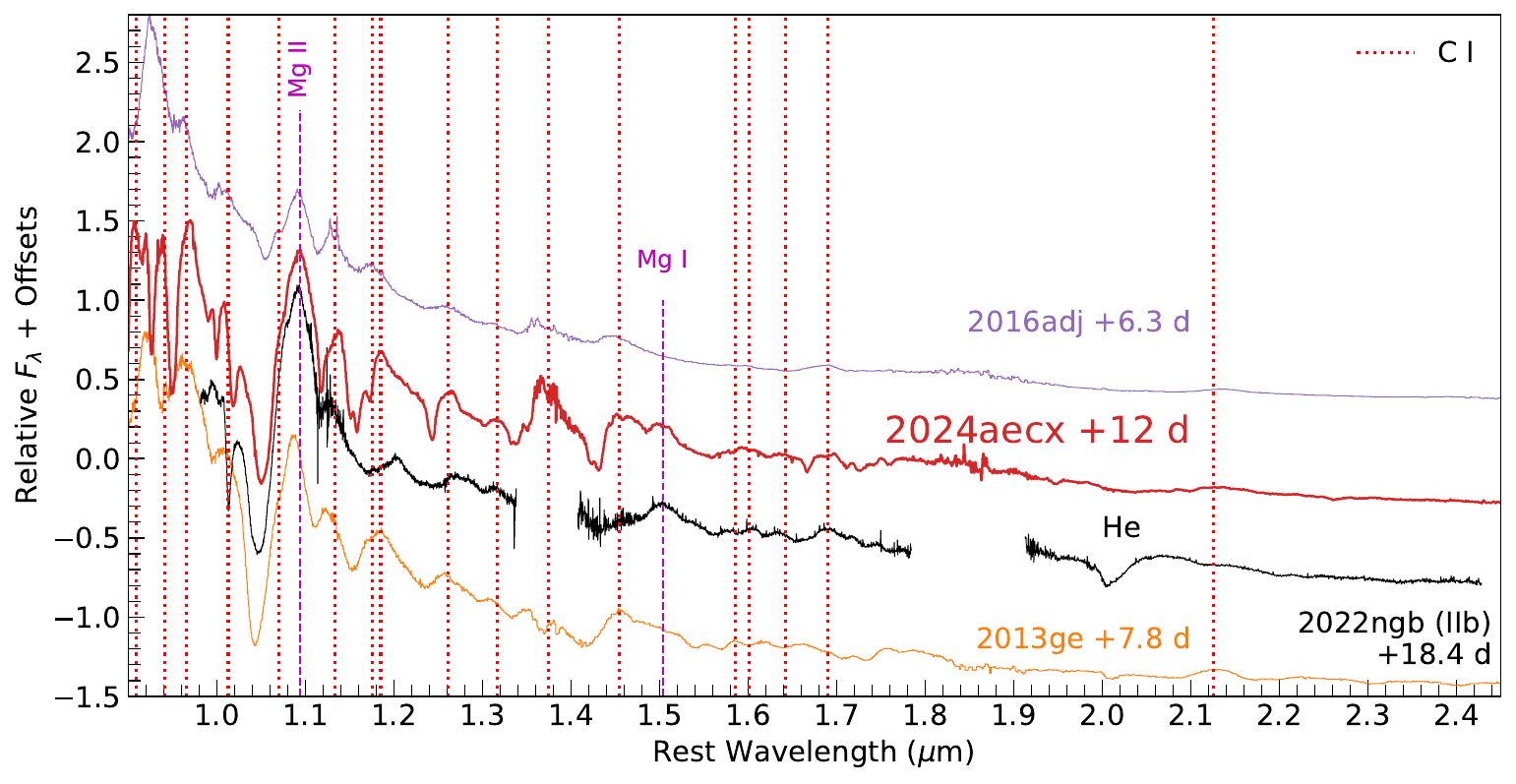}
    \includegraphics[width=\linewidth]{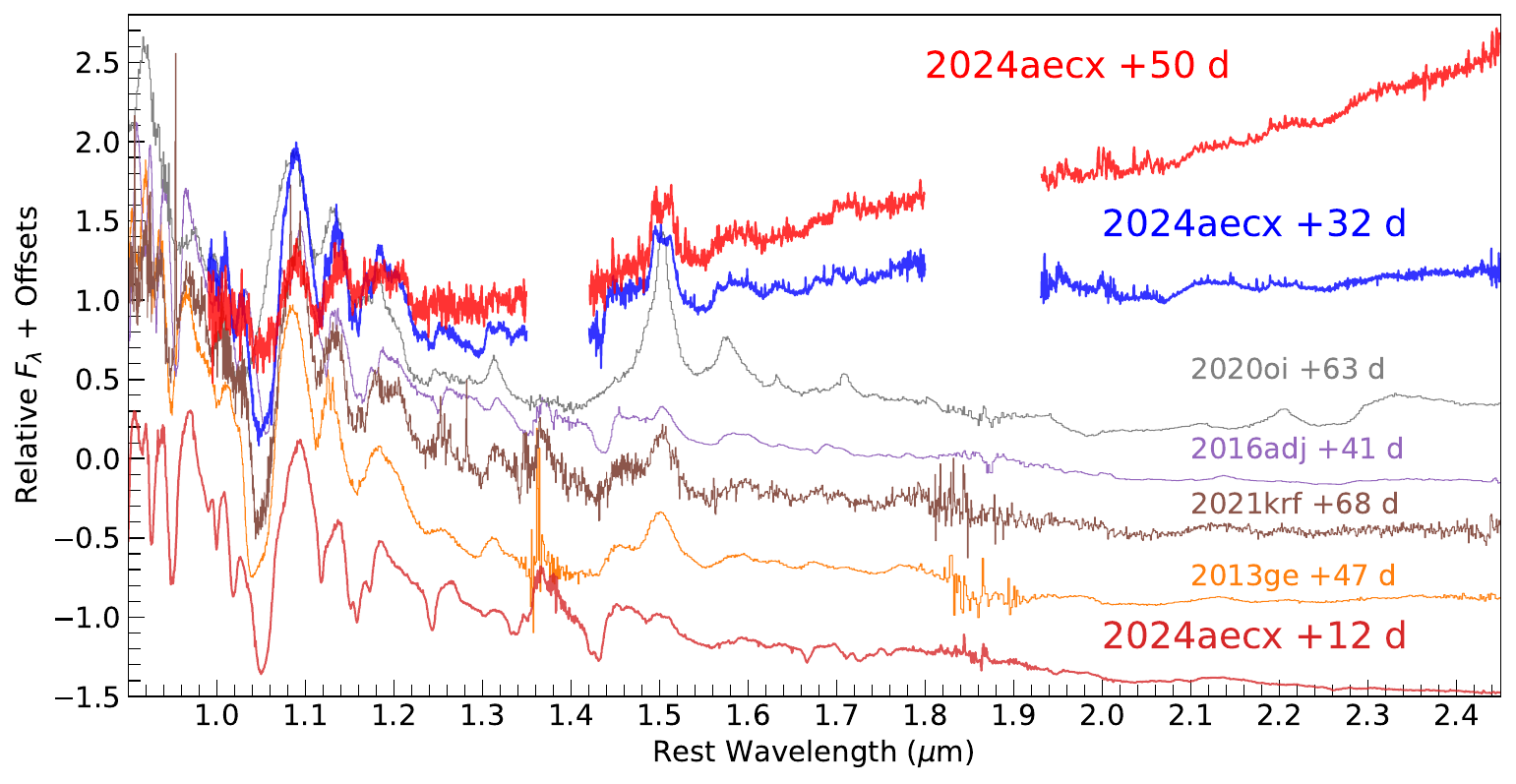}
    \caption{
    \textbf{Top: }NIR spectrum of SN\,2024aecx at 12 days post peak compared with those of a typical SN Ic 2013ge at 7.8 days \citep{Shahbandeh2022}, C-rich SN Ic 2016adj at 6.3 days \citep{Stritzinger2024}, and a helium-rich SN IIb 2022ngb at 18 days \citep{Tinyanont2024}. 
    \ion{C}{1} lines are plotted in dotted red lines. Other lines present are \ion{Mg}{1} 1.5003 $\mu$m and \ion{Mg}{2} 1.0938 $\mu$m, plotted in dashed magenta lines. 
    The conspicuous helium absorption associated with the 2.059 $\mu$m line seen in SN\,2022ngb is absent in SN\,2024aecx. 
    \textbf{Bottom: }NIR spectra of SN\,2024aecx representative of different phases, compared with the spectra of a typical SN Ic 2013ge \citep{Shahbandeh2022}; a strongly reddened SN Ic 2016adj with early CO formation \citep{Stritzinger2024}; the CO- and dust-forming SNe\,2020oi \citep{Rho2021} and 2021krf \citep{Ravi2023}. This comparison highlights the extreme nature of the IR excess in SN\,2024aecx. 
    }
    \label{fig:spec}
\end{figure*}

\section{Host galaxy extinction}\label{sec:extinction}
SN\,2024aecx experiences significant dust extinction from its host galaxy, in addition to the Milky Way (MW) extinction of $E(B-V)_{\rm MW} = 0.05$ mag \citep{Schlafly2011}.
This is demonstrated by the strong and saturated \ion{Na}{1} doublet absorption at the host redshift seen in the public TNS spectrum from 2024 December 19 \citep{Andrews2024b}. 
While our primary observations are in the NIR and our results are not heavily affected by interstellar dust extinction, we need an estimate of the host extinction to rule out the possibility of extreme reddening as seen in, e.g., the Type Ic SN\,2016adj ($A_V \sim 7$ mag; \citealp{Stritzinger2024}).

We downloaded a public optical spectrum from the TNS classification report, obtained with the Gemini Multi-Object Spectrograph North on 2024 December 19, about 2 days after the SN discovery \citep{Andrews2024b}. 
The spectrum shows clear \ion{Na}{1} doublet absorption both from the MW and from the host galaxy. 
The lines are partially resolved. 
We simultaneously fit the continuum and both \ion{Na}{1} doublet from MW and the host using Gaussian line profiles, and measured equivalent widths (EW) of 0.87 \AA\ from the MW and 2.63 \AA\ from the the host galaxy. 
This EW is outside the calibrated range in \citet{Poznanski2012}. 
If we assume that $A_V$ scales linearly with the \ion{Na}{1} EW (e.g, \citealp{Stritzinger2018}), then we estimate $A_{V, \rm host} \sim 0.48$ based on the MW extinction. 
We note that using the \ion{Na}{1} EW to infer the host extinction results in a large scatter \citep[e.g.,][]{Phillips2013, Stritzinger2018}, and we did not use this estimate further.

To provide a better host extinction estimate, we compare the observed optical color evolution of SN\,2024aecx to the color templates of SESNe observed by the CSP-I \citep{Stritzinger2018}, which are constructed from the subsample of SESNe with minimal host extinction. 
We use the \texttt{scipy.optimize} package to minimize the least squares between the templates and the observed colors by computing the $B-V$, $V-r$, and $r-i$ colors and applying the reddening law of \citet{Fitzpatrick1999} leaving both $R_V$ and $A_V$ as free parameters to match the CSP SNe IIb, Ib, and Ic templates.

\begin{figure*}
    \centering
    \includegraphics[width=\linewidth]{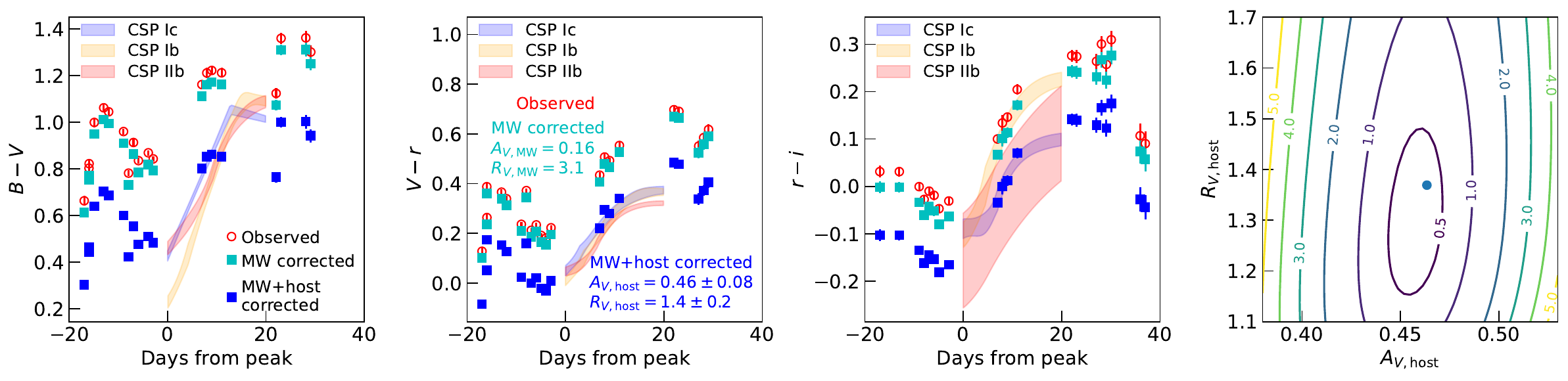} \hfill
    \includegraphics[width=0.8\linewidth]{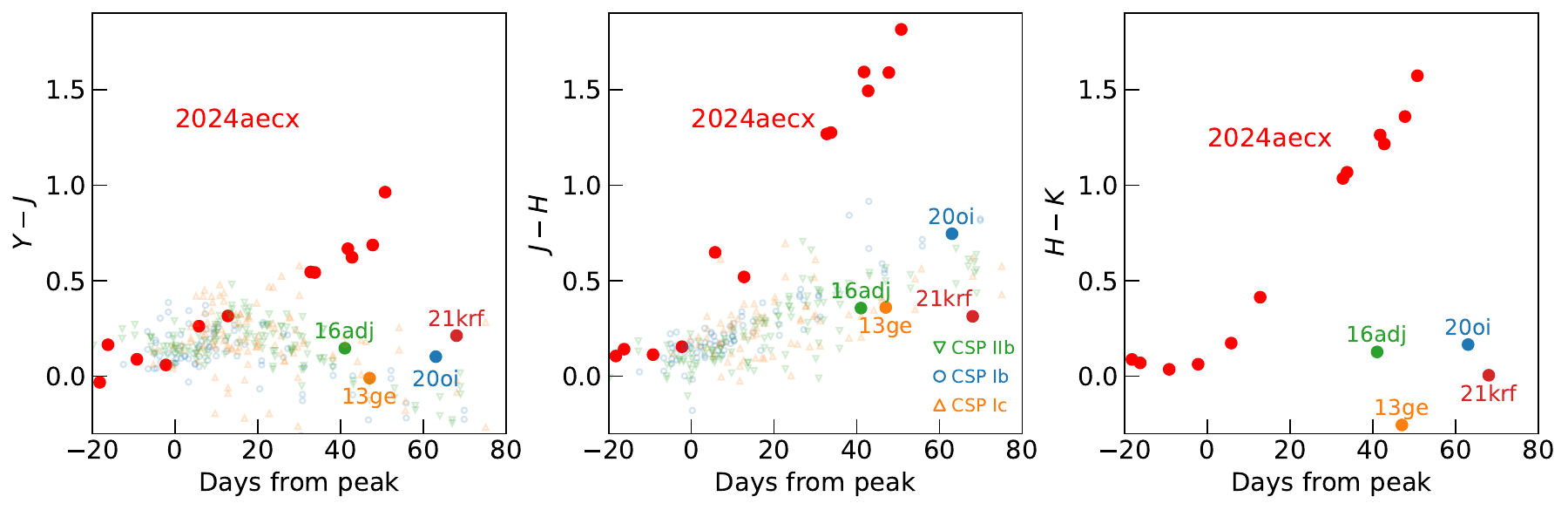}
    \caption{
    \textbf{Top:} color evolution of SN\,2024aecx $B-V$ (left), $V-r$ (middle left), and $r-i$ (middle right). Unfilled red circles mark the observed color. Filled cyan squares mark the colors corrected for MW extinction with $A_{V, \rm MW} = 0.16$ mag and $R_{V, \rm MW} = 3.1$. Filled blue squares mark the colors corrected for both MW and host extinction with {$A_{V, \rm host} = 0.46 \pm 0.08$ mag and $R_{V, \rm host} = 1.4 \pm 0.2$.} Shaded bands mark the range of color templates for SNe IIb, Ib, and Ic from CSP-I \citep{Stritzinger2018}. The Ic template is used to determine the host extinction parameters. {The normalized chi-square contour of the CSP-I Ic template fitting used to find the best values for $A_{V, \rm host}$ and $R_{V, \rm host}$ is shown in the rightmost panel. }
    \textbf{Bottom:} synthetic NIR colors evolution of SN\,2024aecx: $Y-J$ (left), $J-H$ (middle), and $H-K$ (right), compared with those of comparison events SNe\,2013ge, 2016adj, 2020oi, and 2021krf. The synthetic colors are computed from spectra presented in Figure~\ref{fig:spec}. 
    Transparent {downward-pointing triangles, circles, and upward-pointing triangles} show photometric colors of SNe { IIb, Ib, and Ic}, respectively, from CSP~I \citep{Stritzinger2018phot}. Note that CSP~I does not have $K$-band observations. 
    }
    \label{fig:synth_NIR_color}
\end{figure*}

We find that the optical color evolution of SN\,2024aecx is not compatible with that of SNe Ib. 
After correcting for the minimal MW extinction, the $r-i$ color already matches well with the Ib template, while there are still significant $B-V$ and $V-r$ color excesses. 
To correct this, the fit requires a very steep reddening curve with $R_V = 0.2\pm0.1$ (and $A_V = 0.1\pm0.1$ mag), well outside of the range found by \cite{Stritzinger2018} in the CSP-I sample.
The IIb template gives more reasonable parameters with $A_V = 0.6\pm0.4$ mag and $R_V = 1.4\pm0.3$. 
As discussed below, our NIR spectra strongly reject types IIb or Ib classifications due to the lack of strong helium features seen in other KITS SNe Ib and IIb \citep{Tinyanont2024}. 
Thus, we use the Ic template to determine the final extinction parameters and find that the best-fit parameters are $A_{V, \rm \ host} = 0.46 \pm 0.08$ mag and $R_{V, \rm \ host} = 1.4 \pm 0.2$ for the host galaxy, where the uncertainties come from the range of intrinsic colors provided by the CSP-I template. 
{We note that the uncertainties are likely underestimated due to the limited number of SNe Ic in the \citet{Stritzinger2018} sample, as discussed by \citet{Rodriguez2023}.}
The evolution of the $B-V$, $V-r$, and $r-i$ colors as observed, with only MW extinction correction, and with the MW plus host correction are shown in Figure~\ref{fig:synth_NIR_color} (top), along with the SNe IIb, Ib, and Ic templates from CSP-I \citep{Stritzinger2018}. 
We note here that the inferred $R_{V, \rm \ host}$ is low compared with other SNe Ic in CSP-I, which tend to prefer higher $R_{V, \rm \ host}$, and is more in line with their one SN IIb with sufficient extinction to determine $R_{V, \rm \ host}$ (SN\,2006T).
While $A_V$ inferred here is lower than what is presented in \citet{Zou2025} and \citet{Xi2025}, our color excess $E(B-V)=A_V/R_V = 0.32$ mag is consistent with their results, as they fixed $R_V = 3.1$ in their analyses.
We note that reddening minimally affects our NIR spectral analyses.

\section{Spectral line evolution}\label{sec:spec_evo}

\subsection{Type Ic classification}

\begin{figure*}
    \centering
    \includegraphics[width=\linewidth]{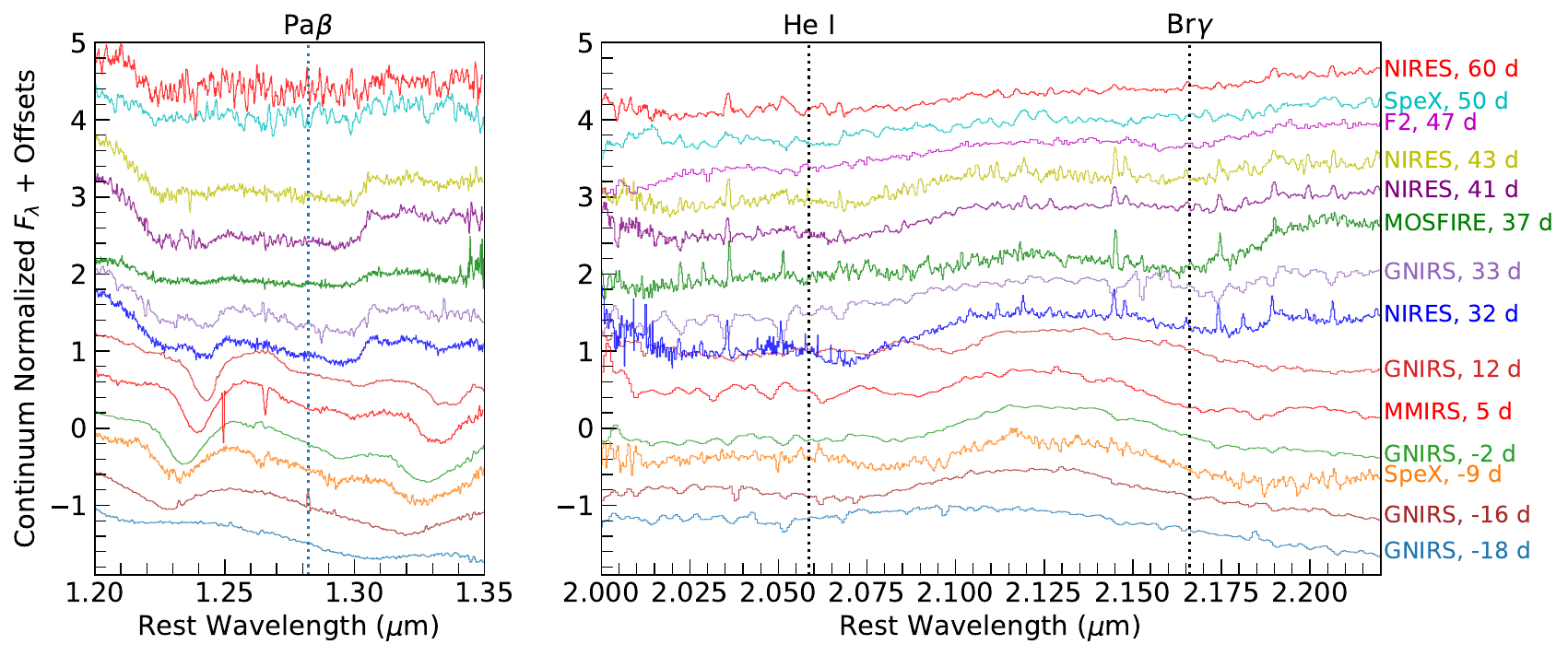}
    \caption{NIR Spectra of SN\,2024aecx at regions around  Pa$\beta$ (left) and the \ion{He}{1} 2.059 $\mu$m and Br$\gamma$ (right). We do not detect hydrogen or uncontaminated helium lines in the NIR at any phase.}
    \label{fig:no_H_He}
\end{figure*}

We do not detect any sign of hydrogen or an unambiguous sign of helium in our spectra. 
Figure~\ref{fig:no_H_He} shows the spectra around isolated NIR hydrogen and helium lines, namely, \ion{He}{1} 2.059 $\mu$m, Br$\gamma$, and Pa$\beta$. 
The uncontaminated \ion{He}{1} 2.059 $\mu$m line is not detected at any phase, in contrast to SNe IIb and Ib from KITS \citep{Tinyanont2024} and CSP~II \citep{Shahbandeh2022}. 
The spectrum of SN IIb 2022ngb is shown in Figure~\ref{fig:spec} (top) for comparison.
SN\,2024aecx has the same absorption feature identified as Pa$\delta$ in SN\,2016adj \citep{Stritzinger2024}, but unlike in that object, SN\,2024aecx never shows the isolated and stronger Pa$\beta$ feature at any phase. 
We prefer the \ion{C}{1} 1.0127 $\mu$m identification, as discussed above.

\citet{Zou2025} presented SN\,2024aecx as a Type IIb with spectral features they identified as \ion{He}{1}~5876~\AA, H$\alpha$, and \ion{He}{1}~7065~\AA. 
However, given the strength of the \ion{C}{1} lines in the IR, we suggest that the lines are due to \ion{Na}{1}~D, and \ion{C}{2}~$\lambda\lambda$6580, 7234, similar to Type Ic SNe\,2007gr \citep{Valenti2008} and 2016adj \citep{Stritzinger2024}. 
Further analysis of optical spectra will be presented in detail in Andrews et al. (2026, in preparation). 

{In addition to the lack of hydrogen and helium in the NIR, we use \texttt{Next Generation SuperFit} \citep{Howell2005, Goldwasser2022} to verify the classification of the public GMOS spectrum from \citet{Andrews2024b}. 
The spectrum is well matched to those of SNe Ic 1994I, 2007gr, and 2014L, in agreement with \citet{Andrews2024b}. 
There is no SN IIb match in the top ten results, which is a strong case against the IIb classification presented in \citet{Zou2025, Xi2025}. 
}

\subsection{Carbon}
Early spectra of SN\,2024aecx ($<$12 days post peak) show a typical hot (a few thousand kelvins) blackbody continuum dominated by \ion{C}{1} absorptions. 
Figure~\ref{fig:spec} (top) compares the spectrum at day 12 with the spectra of Type Ib/c SN\,2013ge \citep{Shahbandeh2022} and Type Ic 2016adj \citep{Stritzinger2024} from similar epochs.
The continuum of SN\,2024aecx is similar to other SNe Ic at this phase. 
In Figure~\ref{fig:ir_spec}, we mark the absorption troughs from the \ion{C}{1} lines at 0.9086, 0.9406, 0.9623, 1.0127, 1.0695, 1.1330, 1.1848, 1.2614, 1.3164, 1.3743, 1.4540, and 1.6890 $\mu$m, most of which are identified in SN\,2016adj \citep{Stritzinger2024}. 
The absorption minimum evolves from around $-13{,}000$ to $-4000\, \rm km\,s^{-1}$ in this plot.
In addition, we identify the weaker \ion{C}{1} 1.5857, 1.6009, 1.6423, and 2.1259 $\mu$m lines, plotted in Figure~\ref{fig:spec} (top).

The \ion{C}{1} absorptions diminish once the IR excess emerges (Section~\ref{sec:ir_excess}), and disappear by $\sim$40 day post peak. 
There are no strong \ion{C}{1} emission at late times, in contrast with calcium, magnesium, and oxygen. 
The high absorption velocity at early times and their disappearance in the nebular phase suggest that carbon is rich in the outer part of the ejecta.

\subsection{Calcium}
\begin{figure}
    \centering
    \includegraphics[width=\linewidth]{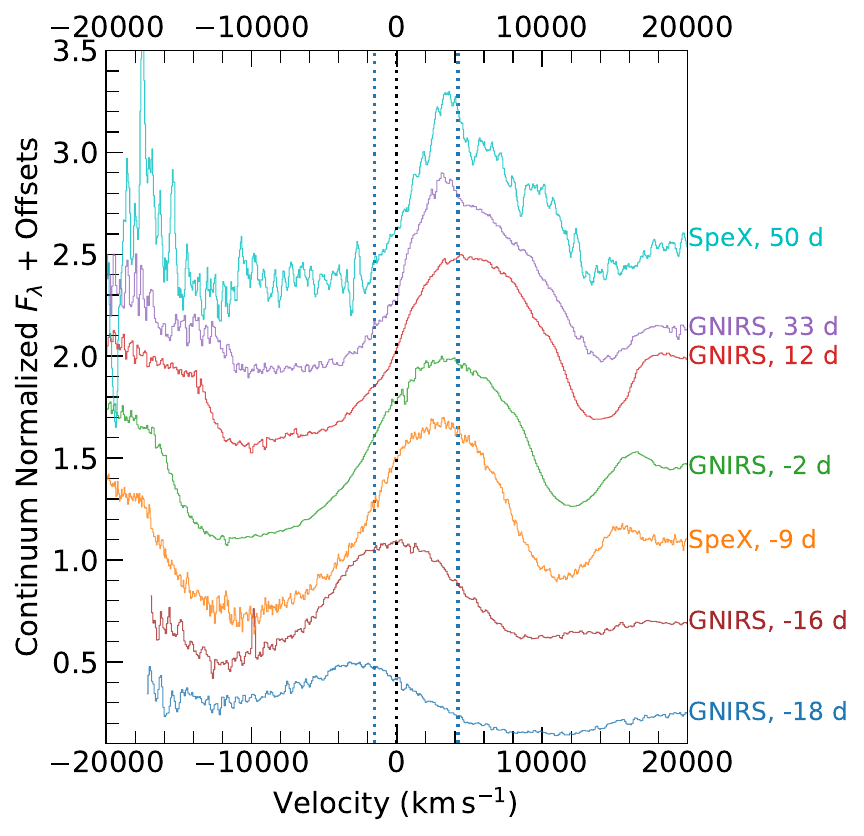}
    \caption{Evolution of the \ion{Ca}{2} triplet from $-18$ to 50 days from peak. Only GNIRS and SpeX cover this line. The velocity is calculated from the 8544 \AA\ line, with the 8500 and 8664 \AA\ lines marked.  
    }
    \label{fig:Ca_triplet}
\end{figure}

Some NIR spectra cover the \ion{Ca}{2} triplet ($\lambda\lambda\lambda$ 8500, 8544, 8664 \AA), which is the strongest line for this SN during its nebular phase. 
Figure~\ref{fig:Ca_triplet} shows the evolution of this line in  velocity space, calculated from the 8544 \AA\ line. 
The emission peaks transition from the 8500 and 8544 \AA\ lines to the 8664 \AA\ line at around peak as the ejecta cool. 
The P-Cygni absorption is observed in all epochs, with the absorption trough at around $-12{,}000 \ \rm km\,s^{-1}$ at the first epoch. 
The absorption is no longer clear by 50 days, indicating that the ejecta are becoming optically thin.
The evolution of the \ion{Ca}{2} triplet is similar to typical SESNe \citep[e.g.,][]{Hunter2009, Holmbo2023}. 
In addition to the triplet, the \ion{Ca}{2} 1.1839 and 1.1950 $\mu$m lines likely cause the multi-component absorption features in the \ion{C}{1}~1.1848~$\mu$m line.

\subsection{Magnesium}
The \ion{Mg}{1} {1.5033} $\mu$m line is the most prominent and unambiguous feature in the NIR. 
The weaker 1.7108 $\mu$m line is also detected, but not the 1.4878 $\mu$m line.
The 1.5033 $\mu$m line is first detected around 12 days post peak, and becomes strong from 32 days until the end of the spectral sequence at 60 days. 
The line has a boxy shape with some absorption in the middle of the line, or a double-horned shape, unlike any other lines observed in this object. 
The narrow feature near the center of the line might be the residual from skyline subtraction. 
Figure~\ref{fig:1micron} (right) shows the evolution of this line in the velocity space from 12 to 60 days post peak, showing a relatively stable line profile.
The full width at half maximum (FWHM) of the line is $\sim 5000 \ \rm km\,s^{-1}$. 
This velocity is much lower than the width of the \ion{Ca}{2} line shown above, where the redshifted wing calculated from the reddest triplet line extends to $\sim10{,}000 \ \rm km\,s^{-1}$.

While the \ion{Mg}{1} {1.5033} $\mu$m line is common in SESNe at this phase, the boxy profile observed here is unique.
The spectral comparison in Figure~\ref{fig:spec} (bottom) shows normal \ion{Mg}{1} line profiles in the comparison objects. 
Further, boxy \ion{Mg}{1} that is constant for over a month is not observed in any SESNe in the first data release of KITS \citep{Tinyanont2024} or in CSP~II \citep{Shahbandeh2022}. 
We note that while \citet{Shahbandeh2022} discussed some SESNe showing ``double-horned'' emission with peaks coinciding with the \ion{Mg}{1} 1.4878 and 1.5033~$\mu$m lines, the \ion{Mg}{1} feature in SN\,2024aecx does not follow this description. 
If the two sides of the boxy profile are centered on these two \ion{Mg}{1}, both have to be redshifted by $1800\ \rm km\,s^{-1}$, which is unlikely given that we do not observe such a large systematic shift in any other lines.

The boxy line profile indicates a line-forming region that is a thin shell, spanning a narrow velocity range in the ejecta \citep{Chevalier1994}.  
This feature is typically seen in H$\alpha$ of SNe II at late times, indicative of a CSM shell or torus heated by interactions with the SN shock \citep[e.g.,][]{Finn1995, Kotak2009, Shivvers2013, Maeda2015, Weil2020, Dessart2022, Shahbandeh2024, Jacobson-Galan2025}.
Alternatively, this could be a double-horned profile from a disk with a hole geometry \citep{Jerkstrand2017}; however, this geometry is less well motivated as the Mg line is unlikely to be emerging from the CSM. 

A potential origin of the boxy \ion{Mg}{1} line is the ejecta heated by same interactions responsible for the early-time shock cooling emission observed in this SN \citep{Zou2025}.
These materials could form a shell behind the forward shock heated by the reverse shock propagating into the ejecta. 
This scenario would explain the constant velocity of this feature over $\sim$1 month of observations. 
If this is the case, then its line-forming region is the CDS between the forward and reverse shocks that could be ripe for later dust formation.  
Alternatively, the \ion{Mg}{1} profile may arise from mixing or asymmetries in the inner ejecta, in which case, the same profile should arise in forbidden lines observed at later epochs.

\subsection{The 1 $\mu$m complex}
\begin{figure*}
    \centering
    \includegraphics[width=0.49\linewidth]{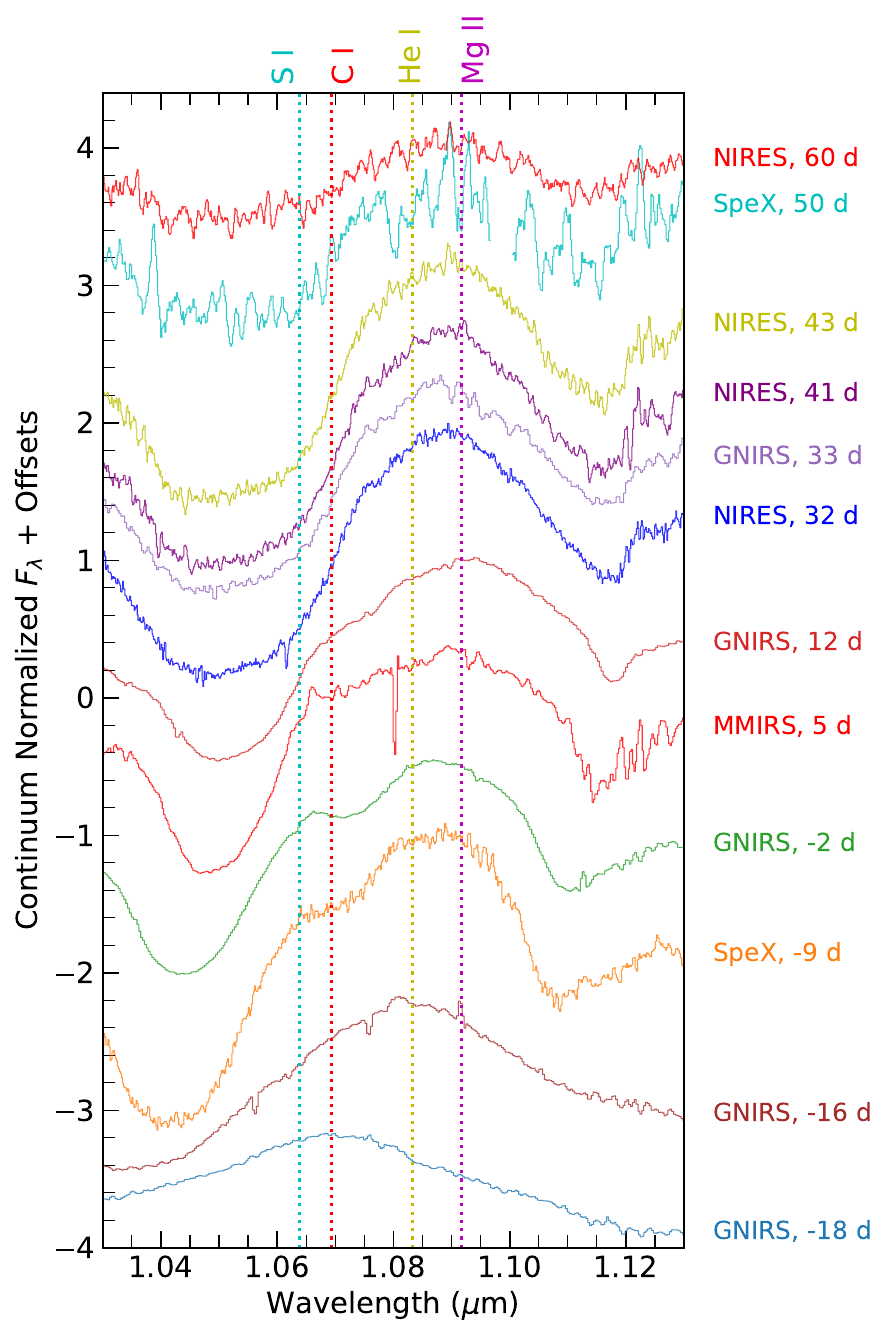} \hfill
        \includegraphics[width=0.49\linewidth]{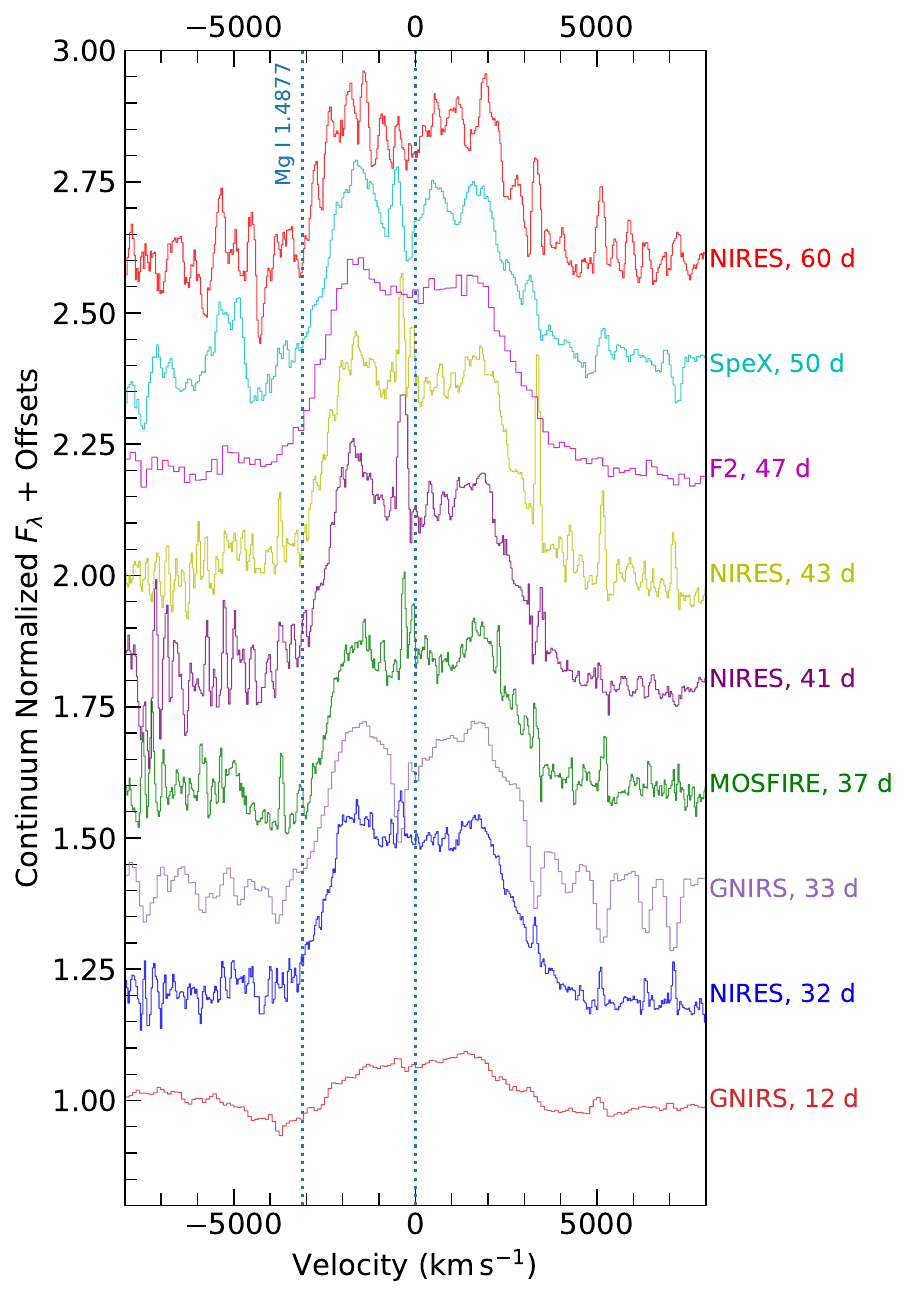}
    \caption{\textbf{Left: }NIR Spectra of SN\,2024aecx between 1.02 and 1.13 $\mu$m, along with possible line identifications of the feature centered around 1.09 $\mu$m.
    \textbf{Right: }evolution of the \ion{Mg}{1} 1.5033 $\mu$m line from 12 to 60 days post peak. The line emerges after peak light.
    The line profile is boxy, with some absorption at $v =0$.
    The FWHM is $\sim$ 5000 $\rm km\,s^{-1}$. 
    The profile remains approximately constant from 32 days post peak.
    }
    \label{fig:1micron}
\end{figure*}

The line complex around 1 $\mu$m is ubiquitous in SESNe; however, the contributions to this feature remain debated \citep[e.g.,][]{Shahbandeh2022}.
Primary lines in this region are \ion{S}{1}~1.0639~$\mu$m, \ion{C}{1}~1.0694~$\mu$m, \ion{He}{1}~1.0833~$\mu$m, and \ion{Mg}{2}~1.0917~$\mu$m. 
Figure~\ref{fig:1micron} shows spectra of SN\,2024aecx covering this region.

Between $-9$ and 12 days from peak, there is a clear shoulder likely due to the \ion{C}{1} 1.0694 $\mu$m line.
This feature fades after 32 days, similar to other \ion{C}{1} lines.
The broad main peak centers between the \ion{He}{1} and the \ion{Mg}{2} lines, including at late times where we expect the line center to be at rest (similar to what is observed in \ion{Ca}{2}). 
Deep P-Cygni absorption can be seen at all epochs, in contrast with the \ion{Mg}{1} 1.5033 $\mu$m line, which only shows weak P-Cygni absorption at 12 days and only emission afterwards.
The lack of the \ion{He}{1} 2.0587 $\mu$m line rules out a large mass of helium. 
Models presented in \citet{Teffs2020} show that as little as 0.2 $M_\odot$ of helium can produce strong \ion{He}{1} 2.0587 $\mu$m at around one week post peak in all cases. 
From this evolution, we propose that a trace amount of \ion{He}{1} contributes significantly to the 1 $\mu$m feature in SN\,2024aecx. 
However, this does not affect its Ic classification, as other SNe Ic also show the same line.

\subsection{Carbon Monoxide}
The CO first overtone band, starting at around 2.3~$\mu$m is tentatively detected in our spectra, starting from around 40 days post peak. 
This feature is ubiquitous in SESNe with late-time observations \citep[e.g.,][]{Hunter2009,Banerjee2018, Rho2021, Shahbandeh2022, Stritzinger2024}.
The epoch at which CO {tentatively} emerges in SN\,2024aecx is in line with typical SESNe. 
Because of the strong NIR continuum and the lack of spectral coverage redward of the CO first overtone band, it is difficult to estimate the strength of the CO emission. 
{As discussed in the next Section, the IR continuum is most likely from IR echoes much further out in the CSM and not from the inner ejecta where CO forms, so they are most likely unrelated.}
Forthcoming JWST data will cover both the CO first overtone, and fundamental bands around 4.6 $\mu$m.

\section{Rapid onset of infrared continuum}\label{sec:ir_excess}
Figure~\ref{fig:ir_spec} shows the NIR spectral sequence of SN\,2024aecx, and Figure~\ref{fig:spec} (bottom) shows the spectra at 32 and 60 days post peak compared with those of SNe Ic from the literature at comparable epochs. 
The comparison events consist of a typical SN Ic 2013ge, well-observed by CSP-II \citep{Shahbandeh2022}; a highly reddened ($A_V \sim 7$ mag) SN Ic 2016adj in the dust lane of Centaurus A \citep{Banerjee2018, Stritzinger2024}; a very nearby SN Ic 2020oi with CO and possible dust formation \citep{Rho2021}; and a peculiar SN Ic 2021krf also with CO and possible hot dust detection \citep{Ravi2023}. 

The spectra of SN\,2024aecx are redder than any of the comparison SNe Ic starting at 32 days post peak.
Figure~\ref{fig:synth_NIR_color} (bottom) shows synthetic NIR colors of SN\,2024aecx and comparison SNe Ic from extinction-corrected spectra at epochs with simultaneous wavelength coverage, along with observed $Y-J$ and $J-H$ colors of SNe { IIb, Ib, and Ic} from CSP~I \citep{Stritzinger2018}.
Before 12 days post peak, the colors of SN\,2024aecx are around 0, similar to SNe Ibc from CSP~I. 
However, starting from 32 days post peak, the colors deviate redward to the maxima of $Y-J = 1.2$ mag, $J-H = 1.8$ mag, and $H-K = 1.6$ mag.
This is 0.6--1.5~mags redder than comparison events at all epochs.
In addition, the NIR colors of SN\,2024aecx are redder than any of the CSP~I SESNe photometric colors \citep{Stritzinger2018}. %
Its extreme NIR colors are in stark contrast with the optical color, which remains very consistent with the SNe Ic population from CSP-I at 40 days post-peak (Figure~\ref{fig:synth_NIR_color}, top).
This behavior cannot be explained by dust in the line of sight and the strong NIR continuum must be from dust local to the SN or an IR echo.

\subsection{Near-infrared continuum properties}
We only use spectra with simultaneous NIR (1--2.4 $\mu$m) coverage (from SpeX, GNIRS, and NIRES) from 32 days post peak onward, which capture the strong NIR continuum. 
As mentioned earlier, the absolute flux scales for SpeX and GNIRS spectra are provided by $z$-band photometry from \citet{Zou2025}, and for NIRES spectra by $K$-band photometry from acquisition images. 
The typical photometric uncertainty is smaller than that of the SN distance. 

\begin{figure*}
    \centering
    \includegraphics[width=0.75\linewidth]{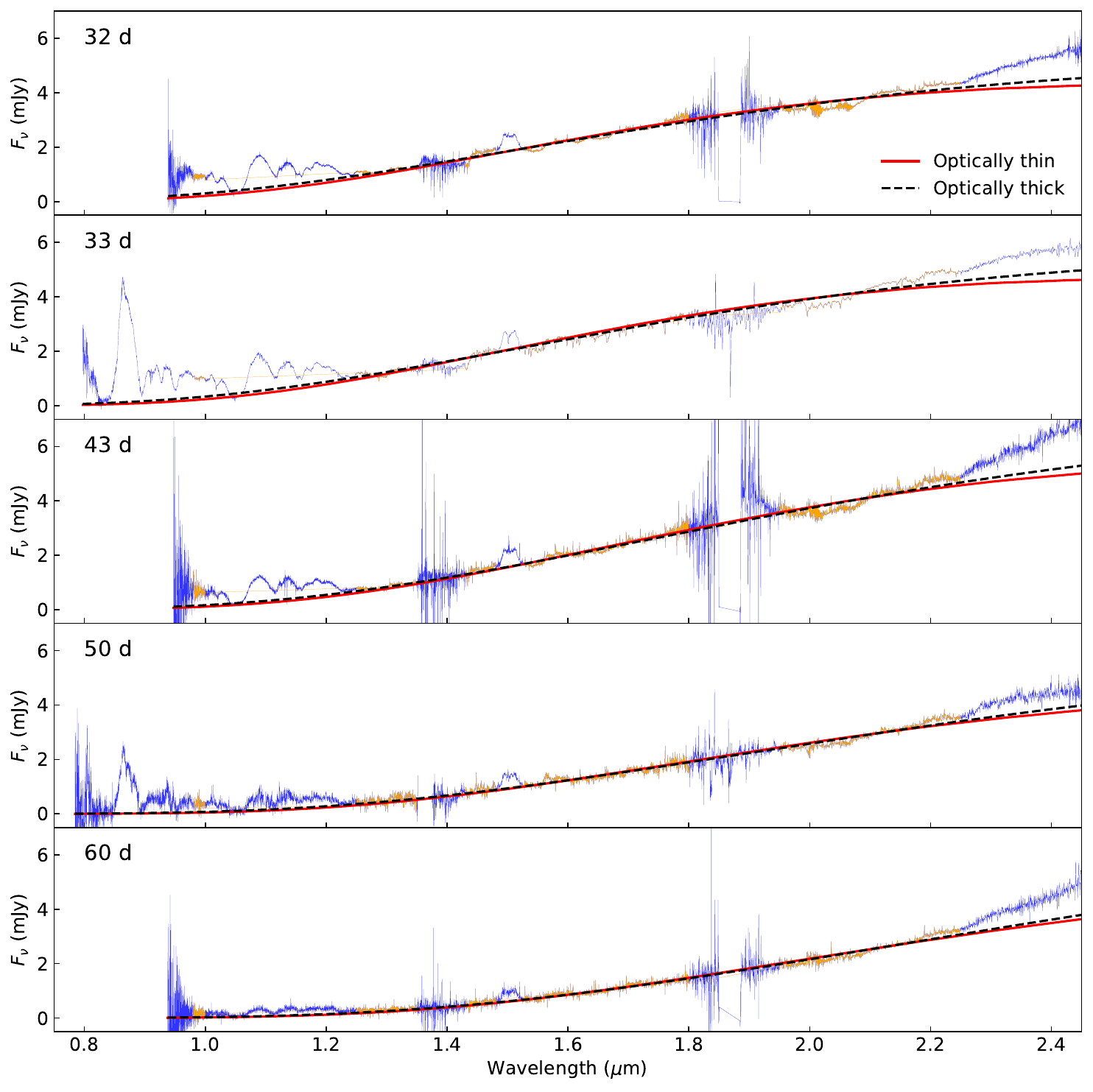}
    \caption{Dust model fit to the observed spectra with absolute flux calibration available. Only the line-free regions plotted in orange are used. The excess beyond 2.3 $\mu$m is likely due to CO emission. The optically thin fit is in solid red, and the optically thick fit is in dashed black. Both scenarios fit the data equally well, from the MCMC results. The dust fit stays below any absorption features (apart from regions near telluric lines), which is expected if the dust continuum is emitted from CSM dust external to line-forming regions. }
    \label{fig:dust_fit}
\end{figure*}

The dust thermal emission has a modified blackbody spectrum depending on the mass, temperature, and grain properties reflected in the dust opacity $\kappa$:
\begin{equation}
    F_{\nu, \rm dust} = \frac{M_{\rm dust} B_{\nu}(T_{\rm dust}) \kappa_{\nu} P_{\rm esc}(\tau)}{d^2}
\end{equation}
The optical depth $\tau$ and the escape probability $P_{\rm esc}$ are calculated assuming spherical geometry, but in the optically thin regime (which we will show is the case later), this assumption does not matter. 
More details of the formulation can be found in \citet{Tinyanont2025}.
We select line-free regions, marked orange in Figure~\ref{fig:dust_fit}, to perform dust model fitting.

In both the optically thick (large $\tau$, blackbody spectrum) and thin cases, we find that only one temperature component is required to fit the bulk of the NIR wavelengths, except for at longer than $\sim$2.25 $\mu$m, where the CO first overtone contributes to the spectrum. 
Without the coverage at longer wavelengths, we cannot distinguish between the optically thick and thin scenarios, nor can we detect colder dust.
NIR data are also insensitive to the grain composition and size; motivated by SN\,2005ip's early spectra showing a lack of silicate feature \citep{Fox2010}, we only consider 0.1 $\mu$m carbonaceous dust with dust opacity data from \citep{Draine1984}.

To fully explore the parameter space, we use \texttt{emcee} to perform Markov Chain Monte Carlo (MCMC) fits to obtain the posterior distribution of dust temperatures, masses, and dust emission region radii. 
The fit accounts for the optical depth, similar to what was done in \citet{Tinyanont2025}.
We use a simple assumption that dust is distributed evenly in a sphere; the real geometry could be asymmetric, leading to a higher density, higher opacity, and even higher underlying dust mass required to explain the observed flux. 
The dust emitting region radius is left as a free parameter, so effectively the optical depth is a free parameter (as it only depends on the dust mass and radius). 
Emission lines are masked before fitting.
The allowed parameter ranges are $0<T<3000$ K, $14 < \log{(r/{\rm cm})} < 20$, and $-7 < \log{(M_{\rm dust}/M_\odot)} < 0$, and we assume uniform prior distributions. 
Figure~\ref{fig:dust_fit} shows the fitting results.

The fits settle in two minima. 
The first minimum represents optically thick dust at a small radius, and the spectrum is a blackbody.
The temperature is generally higher in this scenario, and the mass constraint is a lower limit, as we only see emission from a small fraction of dust grains. 
The radius we get from this scenario is the blackbody radius ($r_{\rm BB}$). 
The second minimum represents optically thin dust at a large, unconstrained radius.
The dust mass is well constrained in this case because we observe all emitted light. 
To better sample the posterior distributions in these two scenarios, we rerun MCMC for the optically thin ($P_{\rm esc} = 1$) and optically thick regimes separately.
The optically thin fit constrains the total dust mass and temperature, while the optically thick fit provides a mass lower limit (5$^{\rm th}$ percentile), temperature, and the blackbody radius. 
In both cases, we integrate the model to get the total luminosity. 
For the optically thin regime, we calculate an equivalent blackbody radius and report it as the size lower limit of the emitting region.
We note that in this case, the dust parameters do not depend on the assumed geometry because we observe emission from all dust grains, and the geometry does not affect the spectra. 
Table~\ref{tab:dust_params} reports these dust parameters.
We note that the reported uncertainties are only statistical. 
Mass, radius, and luminosity measurements are dominated by systematic uncertainties from the distance error; temperatures are more robust, but still subject to uncertainty from flux calibration.

\begin{table*}
\centering
\caption{Dust parameters from MCMC fits, assuming optically thin and thick scenarios}
\label{tab:dust_params}
\begin{tabular}{ll|lllllll}
\toprule
\multicolumn{8}{c}{Optically Thin} \\
\hline
$t_{\rm obs} - t_{\rm peak}$& $t_{\rm obs} - t_{\rm explosion}$ & $T$ & $dT$ & $M$ & $dM$ & $L$ & $dL$ & $r_{\rm BB}$  \\
\hline 
(days) & (days) &(K) & (K) & ($10^{-4}\ M_\odot$) & ($10^{-6}\ M_\odot$) & ($10^{41}\ \rm erg\,s^{-1}$) & ($10^{38}\ \rm erg\,s^{-1}$) & ($10^{16} \, \mathrm{cm}$)  \\
 \hline 
32 & 54& 1126.6 & 0.1 & 0.66 & 0.061 & 1.4 & 2.6 & $>$1.1  \\
33 & 55& 1135.1 & 1.7 & 0.68 & 0.75 & 1.5 & 39 & $>$1.1  \\
43 & 65& 1023.9 & 0.2 & 1.3 & 0.17 & 1.5 & 4.8 & $>$1.4  \\
50 & 72&955.6 & 0.9 & 1.5 & 1.0 & 1.2 & 19 & $>$1.4  \\
60 & 82&869.8 & 0.4 & 2.6 & 1.1 & 1.2 & 15 & $>$1.7  \\
\hline
\multicolumn{8}{c}{Optically Thick} \\
\hline
$t_{\rm obs} - t_{\rm peak}$& $t_{\rm obs} - t_{\rm explosion}$& $T$ & $dT$ & $M$ & $L$ & $dL$ & $r_{\rm BB}$ & $dr_{\rm BB}$ \\
\hline
 (days) & (days) &(K) & (K) & ($10^{-3}\ M_\odot$) & ($10^{41}\ \mathrm{erg\,s^{-1}}$) & ($10^{-38}\ \mathrm{erg\,s^{-1}}$) & ($10^{16} \, \mathrm{cm}$) & ($10^{13} \, \mathrm{cm}$)  \\
 \hline
32 &54& 1583.9 & 0.3 & $>$ 4.7 & 1.7 & 1.3 & 0.62 & 0.29 \\
33 &55& 1587.6 & 3.4 & $>$ 0.85 & 1.9 & 16 & 0.65 & 3.6 \\
43 &65& 1391.8 & 0.3 & $>$ 7.0 & 2.0 & 1.8 & 0.87 & 0.54 \\
50 &72& 1278.4 & 1.5 & $>$ 2.5 & 1.6 & 7.5 & 0.91 & 3.1 \\
60 &82& 1130.5 & 0.8 & $>$ 5.7 & 1.7 & 4.7 & 1.2 & 2.6 \\
\hline
\end{tabular}
\end{table*}

\subsection{Dust parameter evolutions}

\begin{figure*}
    \centering
    \includegraphics[width=\linewidth]{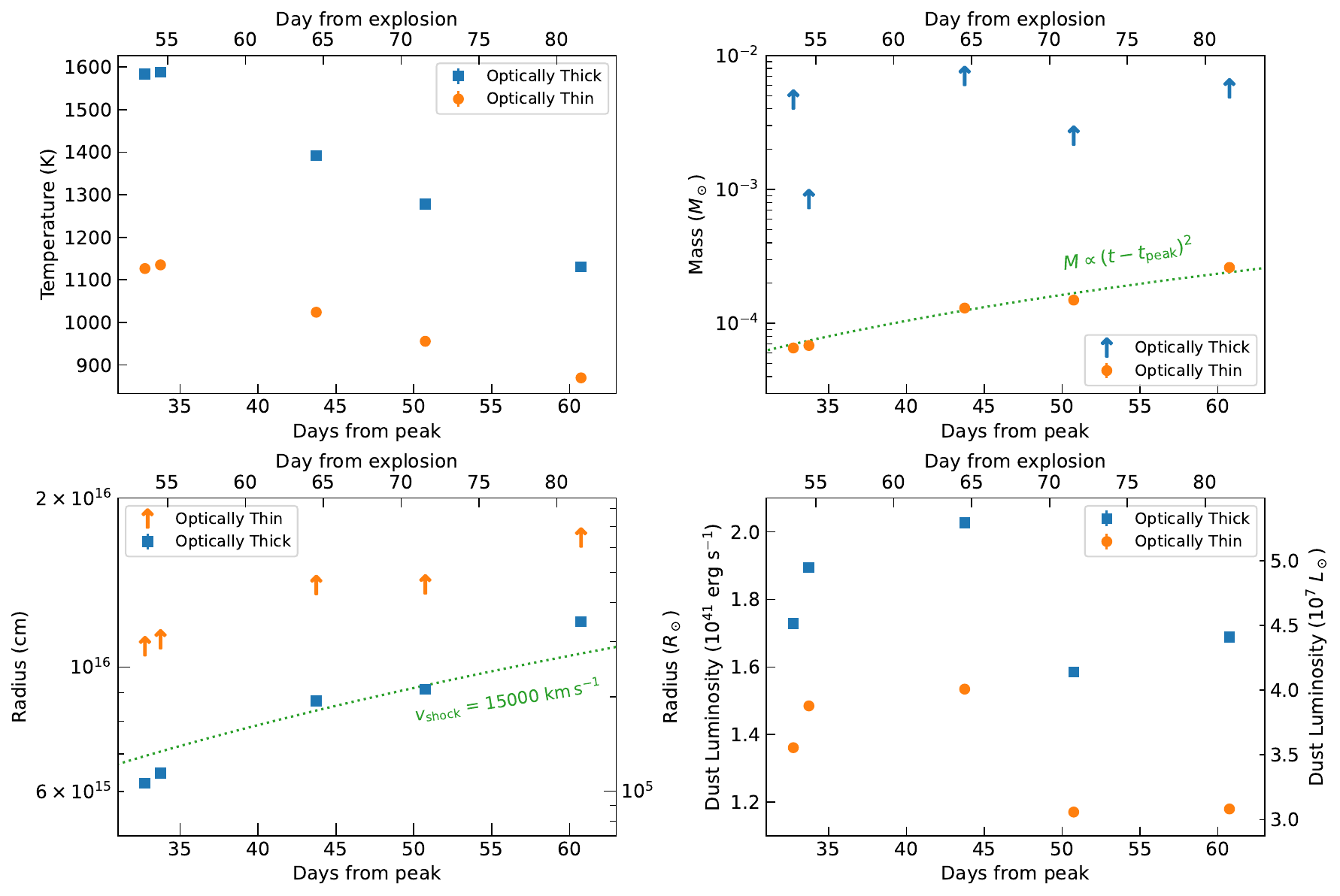}
    \caption{Evolution of dust parameters in SN\,2024aecx, clockwise from top left: temperature, mass, radius, and luminosity.
    Results from the optically thick and thin are shown in different colors and markers. 
    Observation epochs are shown both with respect to the explosion and the peak.
    Lower limits are shown as upward arrows. 
    Statistical uncertainties are plotted, but are small compared to the marker size. 
    In the mass plot, the trend $M\propto (t-t_{\rm peak})^2$ is shown.
    In the radius plot, the shock radius assuming $v_{\rm shock} = 15{,}000 \ \rm km\,s^{-1}$ is shown. 
    }
    \label{fig:dust_evo}
\end{figure*}

Figure~\ref{fig:dust_evo} shows the evolution of the dust mass, temperature, luminosity, and radius for both optically thick and thin cases. 
The temperature declines almost linearly in both cases, with the optically thick dust being a few hundred kelvins hotter.
In the optically thick case, the hottest dust observed is at around 1600 K, which is right at the sublimation temperature. 
The dust mass required to explain the NIR continuum is $\sim 10^{-4} \ M_\odot$ in the optically thin case, and $\gtrsim 10^{-3}\ M_\odot$ for the optically thick case.
The well-constrained optically thin dust mass increases as $M\propto (t-t_{\rm peak})^{2}$. 
The dust-emitting region radius increases monotonically in both cases. 
The shock radius, assuming a shock velocity of $v_{\rm shock} = 15{,}000\ \rm km\,s^{-1}$, is shown for reference. 
The velocity is arbitrarily chosen to pass through the optically thick radius; this velocity is consistent with early-time spectroscopy.
The radius lower limit for the optically thin case is about a factor of 2 larger.
Finally, the dust luminosity obtained by integrating the NIR spectral energy distribution (SED) model is similar in both cases.
The difference arises from the unobserved extra flux at longer wavelengths for the optically thin dust versus the blackbody. 
The luminosity is roughly constant in both cases.

\subsection{Infrared echo as the origin of the infrared excess}

The rapid onset of the NIR excess without strong interactions (due to the lack of additional flux) and the dust parameter evolution indicate that the most likely culprit is an IR echo \citep{Bode1980, Wright1980, Dwek1983}.
In this scenario, the SN light heats up pre-existing dust grains, which subsequently cools by emitting in the IR. 
Similar to a normal light echo, at any given time delay from the light source (e.g., SN shock breakout or peak), we observe emission from an ellipsoid with the SN and the observer at the two foci, which can be approximated as a paraboloid at such a distance. 
This expanding paraboloid could intersect ISM dust grains unrelated to the SN, heating them to IR-emitting temperatures.

The NIR excess in SN\,2024aecx shows high temperature $\sim1000$ K, which declines over time, indicating that the dust temperature is distance dependent and the dust is in the CSM close to the SN. 
In this regime, the dust grains are heated by the SN flux (which could be from the shock breakout, cooling, or the peak), so the grain temperature is distance dependent.
\citet{Fox2010} presents a calculation for dust temperature at different radii provided the SN luminosity in their Equation 13, which can be rearranged to:
\begin{equation}
    r_{\rm dust} = \left( \frac{3\  L_{\rm SN}}{64\ \rho a \sigma_{\rm SB} T_{\rm SN}^4} 
    \frac{\int_0^\infty B_\nu(T_{\rm SN}) Q_{\rm abs}(\nu)\  d\nu}{ \int_0^\infty B_\nu(T_{\rm dust}) \kappa_\nu\  d\nu}
    \right)^{1/2} 
\end{equation}
where $\rho \approx 3\ \rm g\,cm^{-3}$ is the dust bulk density, $a$ is the dust grain radius (assumed to be 0.1 $\mu$m), $\sigma_{\rm SB}$ is the Stefan--Boltzmann constant, $Q_{\rm abs}(\nu)$ is the dust absorption efficiency, and $\kappa_\nu$ is the mass absorption coefficient. 
We note that $Q_{\rm abs}(\nu)$ and $\kappa_\nu$ are related by Equation 3 in \citet{Fox2010}. 
As noted by \citet{Fox2010}, the resulting dust temperature is only weakly dependent on the source temperature because the total dust absorption for the source with different temperatures is not significantly different. 

\begin{figure}
    \centering
    \includegraphics[width=\linewidth]{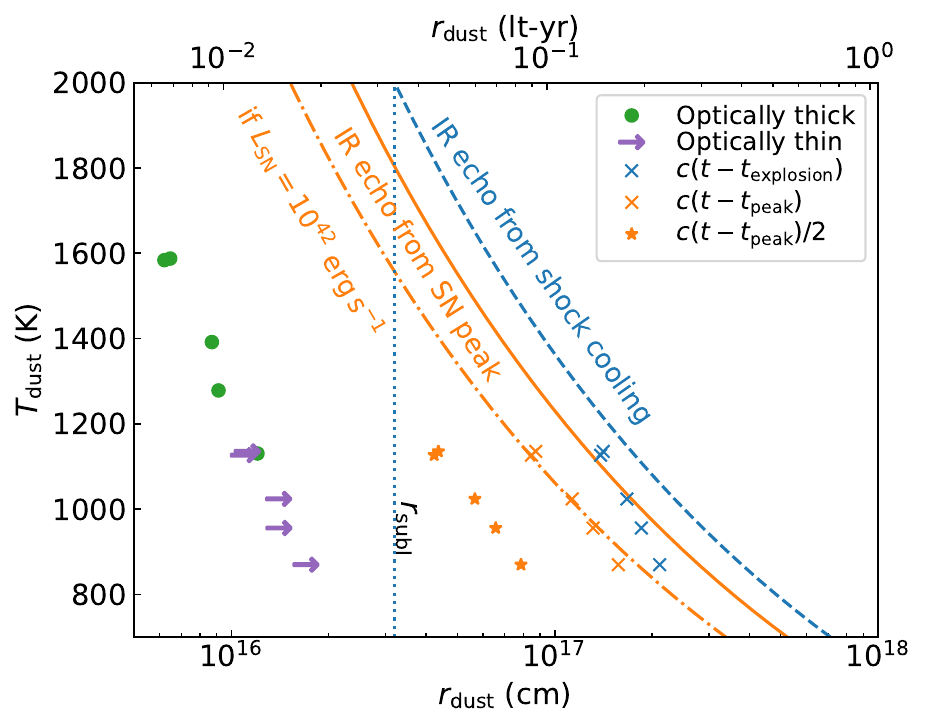}
    \caption{Dust radius and temperature phase space, with SN\,2024aecx dust parameters in the optically thin (lower limits) and thick cases shown. 
    The crosses indicate the light radii of the shock cooling and the SN peak $c(t-t_{\rm explosion/peak})$ at the corresponding epoch of observations for the optically thin dust data points. The focal length of the paraboloid of equal arrival time, $c(t-t_{\rm peak})/2$, is also shown for the SN peak echo scenario.  
    Solid and dashed lines show the dust temperature at different radii if the dust is heated by the shock cooling ($T_{\rm SN} = 10{,}000 \ \rm K$; $L_{\rm SN} = 4 \times 10^{42} \ \rm erg\,s^{-1}$) and SN peak ($T_{\rm SN} = 7000 \ \rm K$; $L_{\rm SN} = 2.5 \times 10^{42} \ \rm erg\,s^{-1}$); both from \citealp{Zou2025, Xi2025}), respectively.
    The dotted line indicates the dust sublimation radius from the shock cooling, assuming the sublimation temperature of 2000 K.
    Any pre-existing dust closer to the SN would have been radiatively destroyed by the shock cooling. 
    }
    \label{fig:ir_echo}
\end{figure}

Figure~\ref{fig:ir_echo} shows dust temperature as a function of distance from the SN, $r_{\rm dust}$ expected from IR echo, assuming two values of $L_{\rm SN}$ from the shock cooling ($L = 4\times 10^{42} \ \rm erg\, s^{-1}$, 10,000 K), the SN peak ($2.5\times 10^{42} \ \rm erg\, s^{-1}$, 7000 K) shown in dashed and solid lines, respectively \citep{Zou2025, Xi2025}. 
We also plot a dash--dotted line for the case that the SN peak is only $1\times 10^{42} \ \rm erg\, s^{-1}$, which is consistent with our observed temperature. 
This could be the case if the host extinction found by \citet{Zou2025, Xi2025} is overestimated. 
{We note that if the distance to the SN is 13.8 Mpc assumed in this paper, as opposed to 11.3 Mpc in \citet{Zou2025, Xi2025}, the luminosity would be even higher, leading to a higher expected IR echo dust temperature. The assumed distance, however, does not affect the observed temperature, only the dust mass. Thus, it does not strongly affect our results.}
We assume $a = 0.1 \  \mu$m carbonaceous dust to be consistent with our dust fit. 
The dotted line shows the dust sublimation radius ($r_{\rm subl}$) out to which all dust grains are heated by the shock cooling peak beyond the assumed sublimation temperature of 2000 K and destroyed. 
The dust parameters from our fit, both the optically thick and thin cases, are shown.
In addition, we plot the light radii $r_{\rm light} = c(t-t_{\rm explosion, peak})$ at the epochs of observations for the optically thin dust, and also the focal length of the echo paraboloid, $c(t-t_{\rm peak}/2).$

For the optically thick case, the inferred dust radius is consistent with the shock velocity. 
The radius is smaller than $10^{16} \ \rm cm$ at all epochs, suggesting that the dust must be newly formed because all pre-existing dust at these radii is destroyed by either the SN shock cooling or peak emission. 
The high observed temperature is consistent with newly condensed dust grains. 
However, the timescale of this dust formation is problematic. 
For the optically thick case, we require at least $10^{-2}\ M_\odot$ of dust as early as 54 days post explosion. 
Typical CCSNe do not form dust until hundreds of days post explosion as the ejecta expand and cool. 
Dust formation mere months post explosion has only been observed in SNe with strong interactions, in which a CDS is formed between the forward and reverse shocks, allowing dust to condense \citep[e.g.,][]{Gall2014, Sarangi2018b, Shahbandeh2025}. 
While SN\,2024aecx shows an early peak likely due to shock cooling from the close-in CSM, it is unclear if the density of this CSM is sufficient to create a CDS massive enough to rapidly form such a large amount of dust due to the lack of strong interaction luminosity in the light curve.
Furthermore, an optically thick dusty CSM would result in an extreme optical extinction to the SN light, which we do not observe. 
Given these caveats, we disfavor the optically thick, new dust formation scenario.
We note that while CDS dust formation has been explored theoretically \citep{Sarangi2022}, no model exists yet for hydrogen-poor SNe interacting with hydrogen-poor CSM, but CDS dust formation has been observed in the hydrogen-poor interacting SN Ibn 2006jc \citep{Smith2008}.

The optically thin case is consistent with an IR echo.
In this regime, the assumed spherical symmetry used in the dust fitting step no longer applies because we see emission from all dust grains (the assumption was in the escape fraction and optical depth calculations).
The radius lower limit derived from a blackbody radius is a vast lower limit for optically thin dust, and the dust location is much larger (otherwise it would have been optically thick).
We show in Figure~\ref{fig:ir_echo} that the observed dust temperature in the optically thin case is roughly consistent with the IR echo if dust is located at the light radius of the echo $r_{\rm light} = c(t-t_{\rm explosion, peak})$.
In other words, either the shock cooling emission or the SN peak can heat dust near the light radius roughly to the observed temperature.

\subsubsection{SN peak as the echo light source}
\begin{figure}
    \centering
    \includegraphics[width=\linewidth]{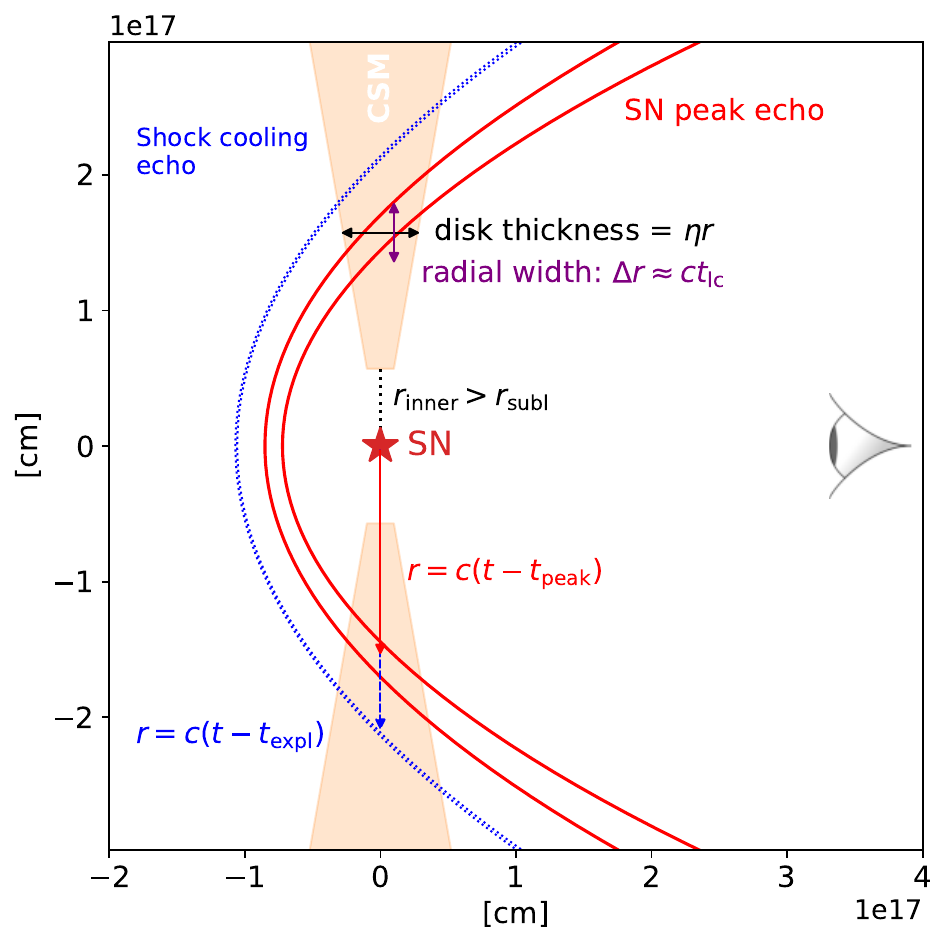}
    \caption{A schematic of the dusty CSM disk responsible for the IR echo. 
    The SN is located at the origin, with the CSM disk shown in cross section aligned with the $y$ axis. The observer is to the right at infinity. 
    The thickness of the disk grows with the radius as $\eta r$, with $\eta = 0.1$ shown here. 
    The disk truncates at an inner radius $r_{\rm inner}$, which is roughly larger than the dust sublimation radius. 
    Two pairs of surfaces of equal arrival time at some observed time $t$ are shown, representing the shock cooling (blue) and the SN peak (red) echo.
    They are parabolas with a focal length of $c(t-t_{\rm expl/peak})/2$, for the shock cooling and the peak echoes.
    In each pair, the two surfaces are separated by the light curve timescale of the source ($t_{\rm lc}$), which is 1 and 10 days for the shock cooling and the SN peak, respectively.  
    The volume of the CSM responsible for the echo observed at time $t$ is where the pair of surfaces intersect the disk, which is along the latus rectum of the parabola at a distance $c(t-t_{\rm expl/peak})$ away from the SN.
    }
    \label{fig:csm_schematic}
\end{figure}

Before we discuss the CSM geometry constraint, we provide an argument that the IR echo is powered by the SN peak and not the shock cooling emission.
At any given epoch $\Delta t$ after the light source, the volume of the CSM dust responsible for the IR echo is between two paraboloids with foci $c (\Delta t \pm t_{\rm lc}/2)/2$ (see Figure~\ref{fig:csm_schematic} for an illustration).
At each epoch, we divide the observed dust mass by the volume of the intersection between this volume and the CSM to get the dust density and then calculate the optical depth.
For the purpose of this calculation, we assume a spherically symmetric CSM to get a lower limit of the dust density (since the dust is the most spread out in this geometry.)
We set the outer radius of the CSM to be a conservative $R_{\rm out} = 5 \times 10^{17} \  \rm cm$, which is where the SN peak could heat the dust to 700 K, below the coolest dust we observe. 
If there is dust beyond, it would only increase the optical depth without showing up in the NIR. 

In this problem, we have a paraboloid whose focus is the SN inside a CSM sphere whose center is also the SN. 
If $F(t) = c t/2$ is the time-dependent focal length of the paraboloid, the distance from the SN to where the two surfaces intersect is $R_{\rm out} - F$.
The volume of the paraboloid up to the point that the two surfaces intersect is $V_{p} = 2 \pi F (R_{\rm out} - F)^2$.
The volume of the top of the CSM sphere above the intersection point (i.e., the paraboloid is outside of the sphere) is $V_{s}(t) = \pi (2F)^2(3R_{\rm out} - 2F)/3$.
We denote $V_{\rm tot} = V_{p} + V_{s}$, and the volume responsible for the IR echo at time $t$ after the light source of timescale $t_{\rm lc}$ can be used to calculate the dust density,
\begin{equation}
\rho_{\rm dust} = \frac{M_{\rm dust, obs}}{V_{\rm tot}(t-t_{\rm lc}/2) - V_{\rm tot}(t+t_{\rm lc}/2)}
\end{equation}
and the optical depth of the entire CSM is $ \tau_{\rm dust} = \pi a^2 \bar{Q} (R_{\rm out} - R_{\rm inner}) n_{\rm dust}$ \citep[Equation 3 in][assuming constant density]{Dwek1983}, where $n_{\rm dust} = \rho_{\rm dust}/(4\pi a^3 \rho_{\rm bulk}/3)$.

Performing this calculation, we find that the optical depth in the shock cooling scenario, where $t_{\rm lc} = 1 \rm \ day$ is at least 1.2--7.9 (values from different epochs), which is inconsistent with the observation that the dust emission is optically thin. 
Also, recall that this optical depth is a lower limit, not accounting for any dust further from the SN that is not heated up to the NIR temperatures. 
These optical depths would also increase if the CSM geometry is not spherically symmetric. 
In contrast, the SN peak scenario with $t_{\rm lc} = 10 \rm \ day$ results in the optical depth lower limits of 0.1--0.8. 
We note that the large range of optical depth calculated here shows that the CSM geometry is actually not a sphere; we will address the geometry in the next subsection.
From this calculation, we favor the SN peak as the primary source of the observed early-time IR echo in SN\,2024aecx. 

We note that the echo light source inferred here is different from what had been observed in IR echoes off of ISM dust in old SNe with shock breakout, most notably in Cas A, which was from a Type IIb SN \citep{Dwek2008, Milisavljevic2024}.
In this regime, stochastic heating by UV individual photons from the shock breakout is sufficient to heat the dust to the observed temperatures, such that the dust temperature is set by the peak of the source's SED, independent of the grain's distance from the SN.
Spitzer observations showed that the luminosity required to power the IR echoes is so high that the source could only be the shock breakout \citep{Dwek2008}.
JWST observations of IR echoes in Cas A resolve extremely thin echo surfaces smaller than 2 lt-days, indicating a short-lived light source, likely the shock breakout or the shock cooling emission (\citealp{Milisavljevic2024}; R. Angulo et al. 2026, in preparation). 
However, this regime of IR echo is unlikely to explain the temperature decline seen in SN\,2024aecx.

\subsubsection{Circumstellar medium geometry}

The evolution of the dust parameters and the observed SED allow us to further constrain the geometry of the echoing CSM. 
The observation hints at a near face-on disk, shown in Figure~\ref{fig:csm_schematic}, for the following reasons. 
We reiterate that our dust fit in the optically thin regime is agnostic to the CSM geometry. 
In a spherical CSM, as discussed in the last subsection, the IR echoing volume spans different distances from the SN, and the echoing dust has a range of temperatures.
So we expect an IR echo with a broader SED composed of blackbodies from a range of temperatures.
However, the NIR emission from SN\,2024aecx can be explained by a single-temperature blackbody, which better matches a CSM geometry where the intersecting volume with the IR echo paraboloids is roughly equidistant from the SN. 
Further, the observed temperature evolution is consistent with the temperature of echoing dust at a distance $c(t-t_{\rm peak})$ (twice the focal length; Figure~\ref{fig:ir_echo}), which is the semi-latus rectum of the paraboloid. 

To better compare different geometries to the observations, we set up a simple simulation with a three-dimensional cartesian grid of $101^3$ cells with the SN in the center. 
We consider four different geometries: a homogeneous sphere, a thick disk where the thickness at a distance $r$ from the SN is $\eta r$, a spherical wind with $\rho \propto r^{-2}$, and a shell with the inner radius equal to $r_{\rm subl}$. 
The outer radius of the sphere and shell is arbitrarily set to $R = 2.5 \times 10^{17} \rm\ cm$, but the results do not depend strongly on it because the SED is dominated by the hotter dust near the vortex of the paraboloid.
For the disk model, we select $\eta = 0.35$ as it is the minimum value for which the dust is not optically thick. 
To better fit the observations, we set the temperature and luminosity of the source to 7000 K and $10^{42} \ \rm erg\,s^{-1}$ (see Figure~\ref{fig:ir_echo}), and compute the dust temperature for each cell.  
Then, we compute the dust emission from each cell assuming the same dust grain size and composition we used earlier for IR spectral fitting, and sum them all to get the SED at each epoch from each geometry.
We fit a single-temperature dust model to this SED to show the extent to which the SED differs from a blackbody, and compare the dust mass and temperature from these fits to our observations (we scale the model so that the mass in the third epoch matches the observation).

\begin{figure}
    \centering
    \includegraphics[width=0.8\linewidth]{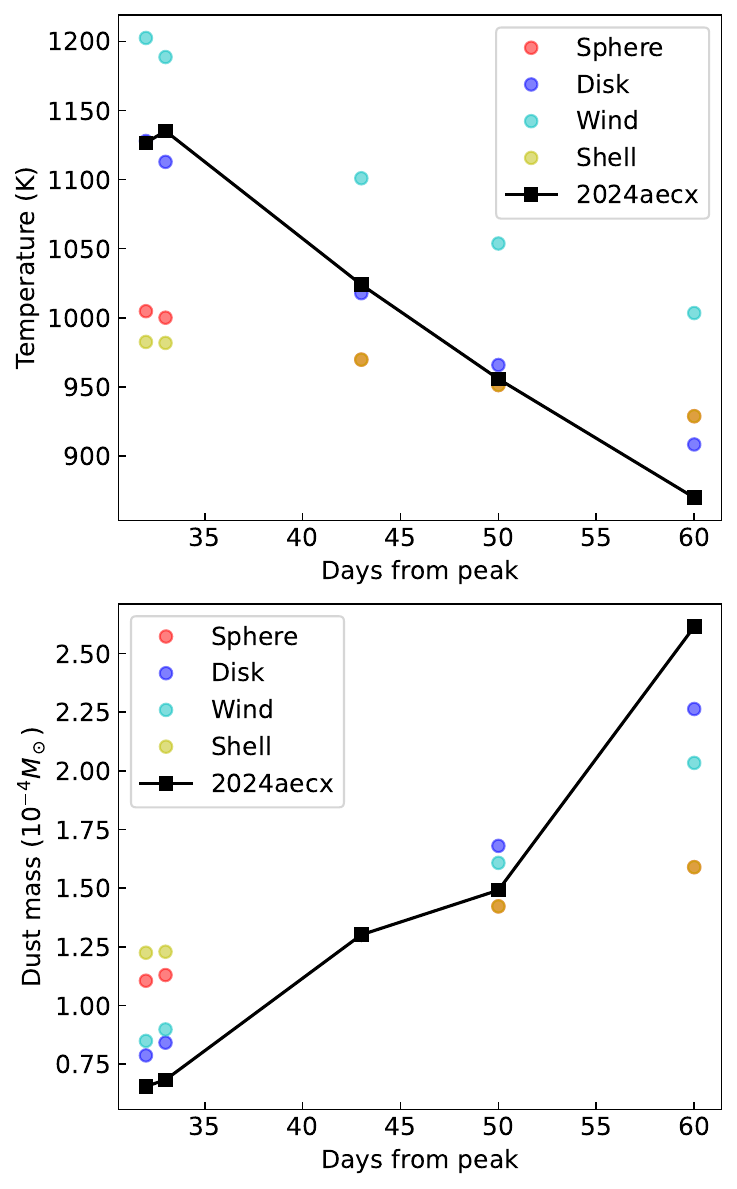}
    \caption{Dust temperature (top) and mass (bottom) evolution from models of IR echo with in the following CSM geometries: homogeneous sphere (red), thick disk (blue), wind with $\rho \propto r^{-2}$ (cyan), and shell with an inner radius of $r_{\rm subl}$ (yellow). The dust parameters of SN\,2024aecx in the optically thin case are shown in black.}
    \label{fig:csm_sim}
\end{figure}

Figure~\ref{fig:csm_sim} compares the dust temperature and mass evolution from different CSM geometries with the observations. 
We first note that the shell and the spherical models give almost identical results because by the first epoch, the echo paraboloid has already swept past $r_{\rm subl}$. 
These models evolve relatively slowly in both dust mass and temperature, because the intersection between the echo paraboloid and the CSM is large and the temperature in the entire volume averages out. 
The wind and disk models produce temperature and mass evolution that are more in tension with the data. 
The temperature evolution of the thick disk model is in good agreement with the data, if we assume a $L = 10^{42} \ \rm erg\, s^{-1}$ light source.
The source luminosity would have to be even lower in the wind scenario. 
Based on this result, we favor a face-on disk, shown in Figure~\ref{fig:csm_schematic}, as the geometry of the CSM responsible for the IR echo in SN\,2024aecx. 

Next, we calculate the IR echo luminosity expected in this scenario. 
Following Equation (2) in \citet{Dwek1983}, we consider a simple case of an optically thin dusty disk with a constant density; the first integral is just the total number of dust grains, which is the total dust mass divided by mass per grain. 
The second integral simplifies to just $L_{\rm SN}/(4\pi r_{\rm dust}^2)$ because the CSM responsible for the IR echo at time $t$ is roughly equidistant from the SN in this geometry. 
We get the IR luminosity fraction, which is just the fraction of the spherical surface area at a distance $r_{\rm dust}$ intercepted by the dust grain (cross section times number of grains):
\begin{equation}
    \frac{L_{\rm IR}}{L_{\rm SN}} = \frac{\pi a^2 \bar{Q} (3M_{\rm dust}/(4 \pi a^3 \rho))}{4 \pi r_{\rm dust}^2} 
\end{equation}
where $\bar{Q} = \int Q_\nu B_{\nu}(T_{\rm SN}) \ d\nu /\int B_{\nu}(T_{\rm SN}) \ d\nu$ is the source averaged absorption efficiency, and $\rho$ is the aforementioned bulk density of the dust grains. 
Performing this calculation, we get $L_{\rm IR} \approx 1.1-2.0 \times 10^{41} \ \rm erg\,s^{-1}$ at all epochs (the range assuming SN peak and shock cooling source), in a good agreement with the observations.

We can also roughly constrain the mass-loss rate of the progenitor star that created this CSM. 
From the dust density calculated previously, if we assume a gas-to-dust mass ratio of 100, we get the gas number density of $n_{\rm CSM} \approx 5\times10^{3} \ \rm cm^3$.
The mass-loss rate that creates the CSM is $\dot{M} \approx 10^{-4} M_\odot\, \rm yr^{-1} (v_{\rm wind}/100 \ \rm km\,s^{-1})$. 
This is much lower than a rate inferred for strongly interacting SESNe, like SN\,2014C \citep{Tinyanont2025}, but that its CSM is (so far) hydrogen-free makes it more similar to, e.g., SN\,2022xxf, whose CSM mass is not constrained in the literature \citep{Kuncarayakti2023}.

Lastly, regardless of the CSM shape, the onset of the NIR excess constrains the inner edge, $r_{\rm inner}$ of the CSM.
With the NIR excess starting between 12--32 days post peak, the inner edge of the CSM is at $(3.1-8.3)\times 10^{16} \ \rm cm$, assuming the SN peak as the echo source.
This is larger than the sublimation radius. 
We also note that the inner edge is similar to that of the CSM around SN\,2014C \citep{Margutti2017}. 
Assuming the shock velocity of $\sim 15{,}000 \ \rm km\,s^{-1}$, consistent with velocities observed in the \ion{C}{1} lines at early times, and $r_{\rm inner}$ from the SN peak echo scenario, the shock should start interacting with the dusty CSM around 240--640 days post explosion creating a different component of dust emission and/or emission lines depending on the CSM composition. 
Follow-up observations of this SN, including those already scheduled with JWST, will further constrain its CSM configuration and potentially detect cooler dust components.

\section{Conclusion}\label{sec:conclusion}
SN\,2024aecx is a hydrogen- and helium-poor Type Ic SN with a rapidly emerging NIR excess starting around one month post peak, unique among SESNe with NIR observations. 
Its peculiar evolution is caught during a regular NIR spectroscopic monitoring by KITS and other surveys, highlighting the necessity of further observations of SNe at these wavelengths.
Its NIR spectra at early times prior to 12 days post peak are similar to those of other SNe Ic, but show strong \ion{C}{1} absorptions with velocities up to $\sim 15{,}000 \ \rm km\,s^{-1}$.
From 32 to 60 days post peak, the NIR continuum shows a strong and evolving excess, likely from dust emission. 
Our analysis shows that the SED evolution could be fit with both optically thick and thin dust models at a single, declining temperature.
We disfavor the optically thick scenario because the large dust mass ($> 10^{-3} \ M_\odot$) required is within the dust sublimation radius, so it could not have been pre existing and has to be all formed within mere months post peak; this is unlikely. 
For the optically thin scenario, the dust properties and evolution can be explained by an IR echo from the SN peak (and not the early luminous but brief shock cooling emission, as it would require a CSM density that is too high and could not have been optically thin).
Our simple simulation shows that a thick face-on disk CSM (Figure~\ref{fig:csm_schematic}) can best explain the dust mass and temperature evolution (Figure~\ref{fig:csm_sim}).
The onset of the NIR excess put a constraint on the inner edge of the disk at $5.7\pm2.6 \times10^{16} \ \rm cm$ from the SN, which is coincidentally similar to the CSM around SN\,2014C \citep{Margutti2017}. 
The SN shock should catch up and start interacting with this CSM around $440\pm200$ days post explosion, and follow-up observations to monitor the onset and the evolution of CSM shock interaction will put a tighter constraint on the CSM geometry. 
Spectroscopy of the interaction will also determine whether this CSM is hydrogen-rich like that around SN\,2014C and other similar objects, or hydrogen-poor like that around SN\,2022xxf. 
Further, IR observations at these epochs will determine whether new dust forms after the CSM is shocked and cooled in the CDS, like what is observed in other interacting SNe \citep[e.g.,][]{Shahbandeh2025, Tinyanont2025}.
Close-in CSM, like what is observed here, is not expected in stars purely stripped by Case B binary mass transfer because the mass-loss episode happens at the end of the progenitor's main sequence, long before its core collapse, and any resulting CSM would have had time to disperse. 
Thus, SESNe with nearby CSM, like SN\,2024aecx, and other similar objects, are key to understanding additional mass-loss process, which may be responsible for stripping helium in SNe Ic. 
Regular IR monitoring of SESNe, even just with photometry, will constrain how common an IR echo is and allow us to observe objects with imminent CSM interactions to better understand the elusive mass-loss mechanisms responsible for stripping their progenitor stars in the last moments of their life. 

\begin{acknowledgments}
This work is supported by the Fundamental Fund of Thailand Science Research and Innovation (TSRI) through the National Astronomical Research Institute of Thailand (Public Organization) (FFB690078/0269).
Time-domain research by the University of Arizona team and D.J.S. is supported by National Science Foundation (NSF) grants 2108032, 2308181, 2407566, and 2432036 and the Heising-Simons Foundation under grant \#2020-1864.
K.A.B. is supported by an LSST-DA Catalyst Fellowship; this publication was thus made possible through the support of Grant 62192 from the John Templeton Foundation to LSST-DA.
L.G. acknowledges financial support from CSIC, MCIN and AEI 10.13039/501100011033 under projects PID2023-151307NB-I00, PIE 20215AT016, and CEX2020-001058-M.
W.J.-G.\ is supported by NASA through Hubble Fellowship grant HSTHF2-51558.001-A awarded by the Space Telescope Science Institute, which is operated for NASA by the Association of Universities for Research in Astronomy, Inc., under contract NAS5-26555.
Supernova research at Rutgers University is supported in part by NSF grant AST-2407567. SWJ gratefully acknowledges support from a Guggenheim Fellowship.
This paper is partially supported by JWST-GO-05451
Time–domain research by the University of California Davis and S. V. is supported by National Science Foundation (NSF) grants AST-2407565.
Some of the data presented herein were obtained at Keck Observatory, which is a private 501(c)3 non-profit organization operated as a scientific partnership among the California Institute of Technology, the University of California, and the National Aeronautics and Space Administration. The Observatory was made possible by the generous financial support of the W. M. Keck Foundation.
Based on observations obtained as part of programs GN-2024B-LP-112 (P.I. Sand $\&$ Andrews) and GS-2025A-Q-122 (P.I. Park) at the international Gemini Observatory, a program of NSF NOIRLab, which is managed by the Association of Universities for Research in Astronomy (AURA) under a cooperative agreement with the U.S. National Science Foundation on behalf of the Gemini Observatory partnership: the U.S. National Science Foundation (United States), National Research Council (Canada), Agencia Nacional de Investigaci\'{o}n y Desarrollo (Chile), Ministerio de Ciencia, Tecnolog\'{i}a e Innovaci\'{o}n (Argentina), Minist\'{e}rio da Ci\^{e}ncia, Tecnologia, Inova\c{c}\~{o}es e Comunica\c{c}\~{o}es (Brazil), and Korea Astronomy and Space Science Institute (Republic of Korea). These observations were processed using DRAGONS (Data Reduction for Astronomy from Gemini Observatory North and South). This work was enabled by observations made from the Gemini North telescope, located within the Maunakea Science Reserve and adjacent to the summit of Maunakea. We are grateful for the privilege of observing the Universe from a place that is unique in both its astronomical quality and its cultural significance. 
Some observations reported here were obtained at the Infrared Telescope Facility, which is operated by the University of Hawaii under contract 80HQTR24DA010 with the National Aeronautics and Space Administration.
Some observations reported here were obtained at the MMT Observatory, a joint facility of the University of Arizona and the Smithsonian Institution. J.R. was partially supported by a NASA ADAP grant (80NSSC23K0749) and JWST-GO-01947.032.

\end{acknowledgments}

\facilities{Keck:II(NIRES), Keck:I(MOSFIRE), Gemini-N(GNIRS), Gemini-S(F2), IRTF(SpeX), MMT(MMIRS), LCO(Swope)}

\software{astropy \citep{astropy:2013, astropy:2018, astropy:2022},
            scipy \citep{2020SciPy-NMeth}, 
            numpy \citep{2020NumPy-Array},
            pypeit \citep{pypeit2020},
            dragons \citep{Labrie2023}, 
            spextool \citep{cushing04},
            xtellcor \citep{vacca03}
          }

\bibliography{2024aecx}{}
\bibliographystyle{aasjournal}

\end{document}